\documentclass{aa}
\usepackage{url}
\usepackage{amsmath}
\usepackage[english]{babel}
\usepackage{mathtools}
\usepackage[varg]{txfonts}

\newcommand{\sectionref}[1]{Sect. \ref{#1}}
\newcommand{\appendixref}[1]{Appendix \ref{#1}}
\newcommand{\figureref}[1]{Fig. \ref{#1}}
\newcommand{\multifigref}[2]{Figs. \ref{#1} - \ref{#2}}
\newcommand{\tableref}[1]{Table \ref{#1}}

\begin{document}

\title{Polarised emission from aligned dust grains in nearby galaxies: predictions from the Auriga simulations}
\titlerunning{Polarised dust emission in Auriga}

\author{B. Vandenbroucke
        \inst{1}
        \and{}
        M. Baes
        \inst{1}
        \and{}
        P. Camps
        \inst{1}
        \and{}
        A. U. Kapoor
        \inst{1}
        \and{}
        D. Barrientos
        \inst{1}
        \and{}
        J.-P. Bernard
        \inst{2}}
\institute{Sterrenkundig Observatorium, Universiteit Gent, Krijgslaan 281,
B-9000 Gent, Belgium
           \email{bert.vandenbroucke@ugent.be}
           \and{}
           Institut de Recherche en Astrophysique et Planetologie (IRAP), Universit\'{e} Paul Sabatier, 9 Av du Colonel Roche, BP 4346, 31028, Toulouse cedex 4, France}

\authorrunning{Vandenbroucke et al.}

\date{Received xxx/Accepted xxx}

\abstract
{
Polarised emission from non-spherical dust grains contains information about the alignment of these dust grains and traces the structure of the interstellar magnetic field.
}
{
In this work, we predict the far-infrared polarisation signal emitted by non-spherical dust grains in nearby galaxies. We determine the angular resolution and sensitivity required to study the magnetic field configuration in these galaxies.
}
{
We post-process a set of Milky Way like galaxies from the Auriga project, assuming a dust mix consisting of spheroidal dust grains that are partially aligned with the model magnetic field. We constrain our dust model using Planck 353~GHz observations of the Milky Way. This model is then extrapolated to shorter wavelengths that cover the peak of interstellar dust emission and to observations of arbitrarily oriented nearby Milky Way like galaxies.
}
{
Assuming an intrinsic linear polarisation fraction that does not vary significantly with wavelength for wavelengths longer than 50 micron, we predict a linear polarisation fraction with a maximum of $10-15~\%$ and a median value of $\approx{}7~\%$ for face-on galaxies and $\approx{}3~\%$ for edge-on galaxies. The polarisation fraction anti-correlates with the line of sight density and with the angular dispersion function which expresses the large scale order of the magnetic field perpendicular to the line of sight. The maximum linear polarisation fraction agrees well with the intrinsic properties of the dust model. The true magnetic field orientation can be traced along low density lines of sight when it is coherent along the line of sight. These results also hold for nearby galaxies, where a coherent magnetic field structure is recovered over a range of different broad bands.
}
{
Polarised emission from non-spherical dust grains accurately traces the large scale structure of the galactic magnetic field in Milky Way like galaxies, with expected maximum linear polarisation fractions of $10-15~\%$. To resolve this maximum, a spatial resolution of at least $1~$kpc is required.
}


\maketitle{}

\section{Introduction}

Large-scale magnetic fields with magnitudes of $\sim{}10~\mu{}$G pervade spiral galaxies like our own Milky Way \citep{2013BeckMagneticFieldsGalaxies, 2015BeckMagneticFieldsSpiral}. From a theoretical point of view, there is evidence that these magnetic fields are seeded from primordial magnetic fields in the early Universe with much smaller amplitudes. These small seed fields are then amplified by a galactic dynamo to the observed strengths \citep{2017PakmorMagneticFieldFormation}. While magnetic fields are hence thought to be unimportant for the early formation of galaxies, their dynamic impact after dynamo amplification can be substantial \citep{2016TabatabaeiEmpiricalRelationLargescale}, especially for the vertical structure of the galactic disc \citep{2012HillVerticalStructureSupernovadriven, 2013GentSupernovaregulatedISMII}.

It is well known that magnetic fields play an important dynamic role on small scales, i.e. the cold atomic and molecular gas where star formation takes place \citep{2005HeilesMillenniumArecibo21, 2007McKeeTheoryStarFormationa}. On larger scales, the dynamic impact of magnetic fields is less clear. While the weak primordial magnetic field is definitely unimportant and is dragged along with the galactic flow, it is unclear whether the amplified magnetic field is still just a passenger, since the magnetic pressure contributes significantly to the total pressure of the ISM \citep{2001FerriereInterstellarEnvironmentOur}. Observations of large-scale magnetic fields in nearby galaxies have revealed spiral magnetic field geometries, even in galaxies that have no clear spiral patterns in surface brightness \citep{2015BeckMagneticFieldsSpiral}. This appears incompatible with a magnetic field that is simply dragged along with the galactic flow and could hint at some decoupling between the ISM and the magnetic field \citep{2013BeckMagneticFieldsGalaxies}. However, this discrepancy could also stem from a difference in geometry between the halo and the thin disc, since magnetic field geometries observed from synchrotron emission sample the galactic halo rather than the galactic disc.

The galactic dynamo itself could consist of a turbulent dynamo, i.e. a transport of magnetic energy from small scales to large scales through an inverse turbulent cascade \citep{2016FederrathMagneticFieldAmplification}, or of a so called $\Omega{}-\alpha{}$ dynamo, where small scale vertical motions are amplified by differential rotation in the galactic disc \citep{1988RuzmaikinMagneticFieldsGalaxies, 2013BeckMagneticFieldsGalaxies}. It is also possible that both mechanisms contribute to magnetic field amplification over time \citep{2014PakmorMagneticFieldsCosmological, 2017PakmorMagneticFieldFormation}. Since these mechanisms distribute the magnetic energy on distinct spatial scales, they will lead to distinct magnetic field geometries and correlate differently to other galactic properties like the star formation rate or gas density distribution \citep{2019AndreProbingColdMagnetised}.

To understand the galactic dynamo and its impact on galaxy evolution, it is hence essential to compare these theoretical models with observations of the large-scale magnetic field geometry in real galaxies. Not all techniques that are used to probe the magnetic field are suitable for this purpose. Observations of polarised CO emission probe dense molecular gas that is likely decoupled from the large-scale magnetic field \citep{2011LiAlignmentMolecularCloud}. Optical and NIR observations probe the imprint of aligned dust grains in extinction of background sources. These techniques are biased towards denser lines of sight, where the extinction polarisation signal is stronger (but not so strong that it makes the background source undetectable), and are sensitive to contamination by scattering \citep{1997WoodModelingPolarizationMaps}. Polarised synchrotron emission is sensitive to the magnetic field strength but depends on the poorly constrained cosmic ray density, which has a scale height that is much larger than that of the dust and gas \citep{2015BeckMagneticFieldsSpiral}. It is therefore likely that synchrotron emission traces the galactic halo rather than the star-forming disc, leading to different geometries for the inferred magnetic field in the Milky Way away from the plane \citep{2015PlanckCollaborationPlanckIntermediateResultsa}. Faraday rotation measurements probe both the direction and strength of the magnetic field, but require sufficiently bright background sources \citep{2019AndreProbingColdMagnetised} and predominantly trace the ionised ISM.

One promising avenue to observe the large-scale structure of magnetic fields is by observing the polarised emission from interstellar dust in the FIR \citep{2000HildebrandPrimerFarInfraredPolarimetry}. In the presence of sufficiently strong magnetic fields, non-spherical dust grains are believed to align perpendicular to the local magnetic field \citep{2015AnderssonInterstellarDustGrain}. These grains emit thermal radiation that is linearly polarised along the longest axis of the grain, leading to an emission signature that is perpendicular to the magnetic field direction in a plane perpendicular to the line of sight. Compared to other methods to trace the magnetic field direction, polarised dust emission has the advantage that it traces cold dust in the star-forming disc. The strength of the polarised emission signal is set by the local properties of the dust and its alignment and does not increase with increasing extinction along the line of sight, as is the case for extinction polarisation measurements. Disadvantages of the method are that the dependence of the strength of the polarisation signal on the strength of the local magnetic field is unknown and that line-of-sight averaging of incoherent polarised emission signals will lead to a decrease of the polarisation signal along dense lines of sight \citep{2008Falceta-GoncalvesStudiesRegularRandom}. This method is hence particularly suitable to trace the magnetic field in regions of relatively low density within the star-forming disc, and is complementary to observations of polarised extinction in the optical and NIR, but without the possible contamination by scattering.

Currently, there is only a limited number of facilities that can observe polarised dust emission near the peak wavelength of thermal dust emission at the angular scales allowing well resolved observations of nearby galaxies, i.e. the HAWC+ polarimeter on SOFIA \citep{2018HarperHAWCFarInfraredCamera} and the SCUBA2 polarimeter on the JCMT \citep{2013HollandSCUBA210000}. Due to their limited sensitivity and small field of view, these instruments are badly suited for large surveys of nearby galaxies. While HAWC+ has indeed been able to trace the magnetic field structure in M51 and NGC~891 \citep{2020JonesHAWCFarinfraredObservations}, these results probe a relatively limited part of the magnetic field geometry compared with similar extinction measurements. Measurements with SCUBA2 \citep{2009MatthewsLegacySCUPOL850a} where limited to extremely bright regions in the centre of M82. Balloon borne experiments such as PILOT \citep{2019MangilliGeometryMagneticField} and BLASTPol \citep{2014GalitzkiBalloonborneLargeAperture} are only suitable for the study of the nearest galaxies. At longer wavelengths, ALMA \citep{2016NagaiALMAScienceVerification} and the NIKA2 instrument on the IRAM 30m telescope \citep{2016ComisHighAngularResolution} can trace FIR and submm thermal dust emission, but these instruments only detect the tail of the thermal emission spectrum, in a spectral range with possibly significant sources of contamination such as spinning dust, free-free and synchrotron emission. The B-BOP polarimeter aboard the cancelled ESA/JAXA mission SPICA \citep{2018RoelfsemaSPICAALargeCryogenic, 2019AndreProbingColdMagnetised} would have provided an alternative window to observe polarised dust emission, although it is unclear whether it would have had sufficient resolution for this purpose \citep{2019AndreProbingColdMagnetised}.

In this work, we provide the first realistic simulations of the polarised emission signature from nearby galaxies in the FIR wavelength range, observed from different inclination angles. We use the Auriga suite of Milky Way type spiral galaxies \citep{2017GrandAurigaProjectProperties} as an input model for radiative transfer simulations using the radiative transfer code \textsc{SKIRT} \citep{2011BaesEfficientThreedimensionalNLTE, 2015CampsSKIRTAdvancedDust, 2020CampsSKIRTRedesigningAdvanced} that self-consistently model the thermal emission by interstellar dust grains. We use an accurate model for the optical properties of spheroidal dust grains, computed using the open-source package \textsc{CosTuuM} \citep{2020VandenbrouckeCosTuuMPolarizedThermal}. We constrain our model based on the Planck observations of the Milky Way at 353~GHz \citep{2015PlanckCollaborationPlanckIntermediateResultsa, 2020PlanckCollaborationPlanck2018Results}, and then use this same model to make predictions for nearby galaxies, observed from different inclination angles and in different broad bands. We compare our predictions with HAWC+ data for M51 and NGC~891, and investigate how the synthetic polarisation signal varies as a function of the instrument resolution and wavelength range. We also make recommendations for future FIR polarimetry missions based on the specifications of the B-BOP instrument.

This paper is organised as follows. In \sectionref{sec:method} we detail our methods. In \sectionref{sec:results}, we present the results for our various post-processing simulations. In \sectionref{sec:discussion}, we analyse what these results tell us about the magnetic field in the Auriga simulations, and we discuss the predicted resolution and sensitivity requirements for future FIR instruments. We present our conclusions in \sectionref{sec:conclusion}.

\section{Method}
\label{sec:method}

\subsection{Intrinsic dust polarisation}
\label{sec:intrinsic_model}

There are two crucial ingredients in the post-processing simulations presented in this work that can affect the synthetic polarisation maps we generate. The first ingredient is the intrinsic polarisation of the dust emission signal, which depends on the assumed dust model. The second ingredient is the geometry of the dust distribution, which is given by the parent simulation. The former determines the strength of the polarised intensity generated in every cell of our model, while the latter determines how this signal is altered through integration along the line of sight.

As in \citet{2020VandenbrouckeCosTuuMPolarizedThermal}, our assumed dust model consists of two components: (partially) aligned silicate grains and non-aligned graphite grains. Additional grain materials could be included, if the necessary material properties required to compute optical properties were available. To restrict the parameter space to the parameters of interest for polarisation, we will assume both components obey an MRN size distribution \citep{1977MathisSizeDistributionInterstellar} with
\begin{equation}
    \Omega{}(a) \sim{} a^{-3.5}, \qquad{} a \in{} [5~{\rm{}nm}, 2~\mu{}{\rm{}m}],
\end{equation}
where $a$ is the size of the dust grains. We will furthermore assume that the silicate grains consist of spheroidal grains with a CDE2 shape distribution \citep{2017DraineShapesInterstellarGrains}. As in \citet{2020VandenbrouckeCosTuuMPolarizedThermal} we will assume that only grains with sizes $a>0.1~\mu{}{\rm{}m}$ are aligned, and we will parameterise the alignment using a single linear alignment parameter, $f_{S,A}$. The composition of the dust mixture itself is governed by a second linear mixture parameter, $f_G$ that encodes the fraction of the dust grains contained in the graphite component. Since the optical properties of non-aligned grains are well approximated by spherical grains, we use the graphite optical properties given by \citet{1984DraineOpticalPropertiesInterstellar}, and set the polarised emission coefficient for these grains to $Q_{\rm{}abs,pol}=0$. The total emission coefficient at a specific wavelength $\lambda{}$, grain size and zenith angle $\theta{}$ is then generally given by
\begin{multline}
Q_{\rm{}abs}(\lambda{}, a, \theta{}) = f_G Q_{{\rm{}abs},G}(\lambda{},a) \\
+(1-f_G)\left[
f_{S,A} Q_{{\rm{}abs},S,A}(\lambda{},a,\theta{}) + (1-f_{S,A}) Q_{{\rm{}abs},S,nA}(\lambda{},a)
\right],
\end{multline}
and similar for the polarised absorption/emission coefficient $Q_{\rm{}abs,pol}$. Note that the only zenith angle dependence in this case is in the emission coefficient for aligned silicate grains, $Q_{{\rm{}abs},S,A}$; the zenith angle dependence of the emission coefficient for non-aligned silicate grains, $Q_{{\rm{}abs},S,nA}$ averages out when taking an ensemble average, while the emission coefficient for the spherical graphite grains, $Q_{{\rm{}abs},G}$, has no zenith angle dependence.

\begin{figure}
    \centering{}
    \includegraphics[width=0.49\textwidth{}]{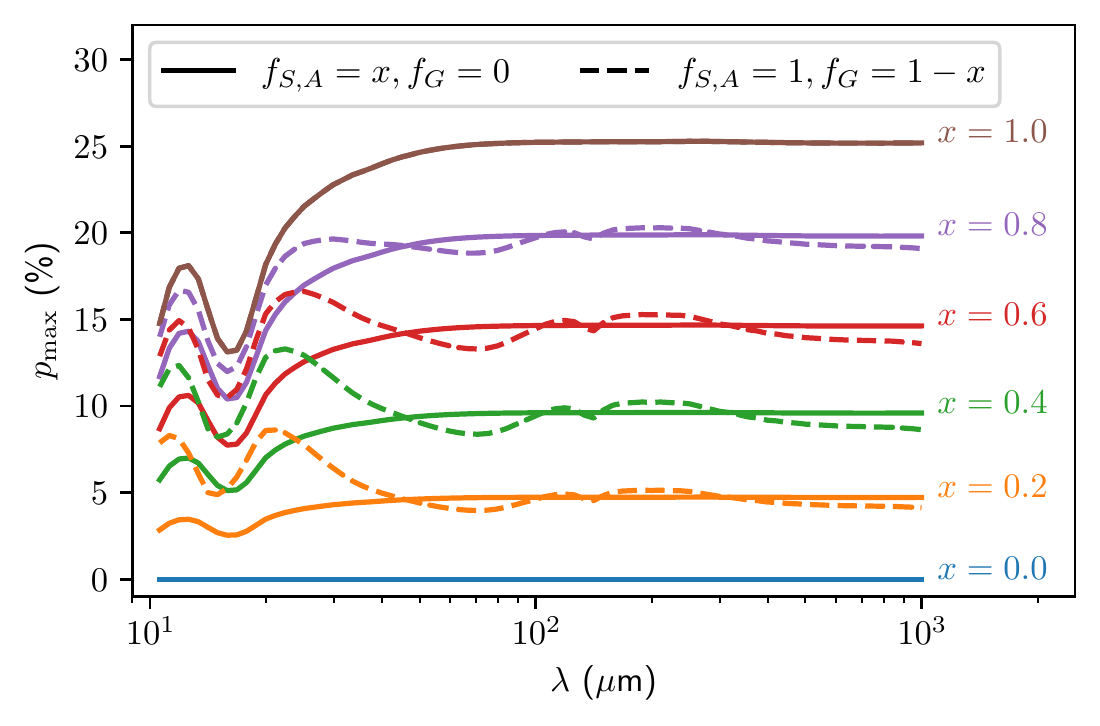}
    \caption{Maximum linear polarisation fraction as a function of wavelength for a dust grain mixture consisting of partially aligned spheroidal silicate grains with a CDE2 shape distribution, and non-aligned spherical graphite grains. The polarisation fraction has been averaged out over an MRN size distribution. The different lines correspond to different grain mixtures: solid lines represent silicate-only mixtures where the alignment fraction $f_{S,A}$ varies with the colour of the line; dashed lines represent a linear mixture of perfectly aligned silicate grains and non-aligned graphite grains.}
    \label{fig:intrinsic_polarisation}
\end{figure}

Under these assumptions, the maximum intrinsic polarisation fraction that can be produced in emission is limited to $p_{\rm{}max} = 25~\%$, as shown in \figureref{fig:intrinsic_polarisation}. This value was obtained by taking the ratio $Q_{\rm{}abs,pol}/Q_{\rm{}abs}$ for our different dust models after integrating the appropriate linear combination of optical properties over the MRN size distribution, and for the emission direction in which polarisation is maximal (perpendicular to the symmetry axis of the spheroidal grains, $\theta{}=\pi{}/2$). Note that this maximum value is consistent with the $p_{\rm{}max}\approx{}22~\%$ observed by Planck in the Milky Way \citep{2020PlanckCollaborationPlanck2018Results}. The values shown only reflect the polarisation fraction that would be observed if the dust emitted at a specific wavelength, and does not factor in the strength of the emission signal that depends on the emission function of the dust, which peaks in a limited wavelength range according to the dust temperature. The maximum intrinsic polarisation is obtained for a perfectly aligned mixture of pure silicate grains, and linearly decreases with both $f_{S,A}$ and $f_G$.

There is a clear degeneracy between the two mixture parameters at long wavelengths, since the intrinsic polarisation for a pure silicate mixture with $f_{S,A}=x$ is very similar to the intrinsic polarisation of a dust mix with $f_G=1-x$ and $f_{S,A}=1$. At shorter wavelengths, it is possible to distinguish between both models due to a decrease in overall emission caused by a decrease in graphite emission. To break the degeneracy, it is necessary to combine observations at $\lambda{}>100~\mu{\rm{}m}$ with observations at $\lambda{}\in{}[20,40]~\mu{}{\rm{}m}$, assuming that it is possible to probe the same dust mixture at both wavelengths in emission, a point that we will investigate in \sectionref{sec:extrapolations}. Since we will limit our analysis to wavelengths of $50~\mu{}$m and more, we will for the remainder of this work assume a silicate only dust mixture with $f_G$ and a single free parameter, $f_{S,A}$.

The shape of the intrinsic polarisation curves at long wavelengths shows that the intrinsic polarisation fraction is only a weak function of wavelength in this regime, consistent with Planck findings at long wavelengths \citep{2015PlanckCollaborationPlanckIntermediateResults} and later work by \citet{2018GuilletDustModelsCompatible}. This is a direct consequence of the wavelength dependence of the material properties used by \textsc{CosTuuM}. Changing the assumed shape distribution or the size limit below which grains are assumed to no longer align only affects the maximum linear polarisation fraction that can be obtained, but does not significantly change the wavelength dependence. The weak wavelength dependence of the intrinsic linear polarisation fraction means that it is theoretically possible to extrapolate Planck results for the Milky Way at long wavelengths to $\lambda{}\in{}[50,450]~\mu{\rm{}m}$, assuming that observations at these wavelengths probe the same dust.

\subsection{Radiative transfer}

The optical properties determined above are used within the radiative transfer code SKIRT to create synthetic observations for various observer positions and bands. The radiative transfer used in these simulations accounts for scattering and absorption of UV and optical light by interstellar dust, and also for the self-consistent heating and thermal emission from this dust \citep{2016CampsFarinfraredDustProperties}. The method used in this work uses a different treatment for thermal dust emission that we briefly outline below.

As in \citet{2016CampsFarinfraredDustProperties}, the SKIRT simulation proceeds in two stages. In a first stage, direct light from stellar sources is propagated throughout the computational domain and used to determine the energy absorption within the interstellar dust. From this, the temperature distribution of the dust grains is calculated, which determines the thermal emission properties for each cell in the domain. In the second stage, the thermal emission spectrum for each cell is in turn propagated to produce the final synthetic maps at long wavelengths. For these simulations, we do not keep track of dust self-absorption, so that this two step approach is sufficient.

In previous works, thermal dust emission was assumed to be isotropic. This translated into an isotropic emission profile for each cell, and also in the use of isotropic weights for peel-off photon packets recorded onto the simulation instruments \citep{2011BaesEfficientThreedimensionalNLTE}. When dust emission originates in aligned spheroidal grains, this approach is no longer valid, since the thermal emission coefficients are now a function of the angle $\theta{}$ between the emitted radiation and the direction of the local magnetic field. This naturally introduces an anisotropy in the emitted radiation. Note that there are two ways to deal with this anisotropy. One could take the ratio of the emission intensity for different directions $\theta{}$ as a difference in weights for photon packets emitted in different directions. Alternatively, one could fix the weight for individual photon packets, and sample directly from the anistropic distribution function. We choose the latter approach.

For each cell, we sample the anisotropic emission profile in a reference frame where the vertical axis coincides with the direction of the local magnetic field. Within this reference frame, the azimuth angle $\phi{}$ of the emitted photon is sampled from a uniform distribution in $[0, 2\pi{}]$. The zenith angle $\theta{}$ is sampled from a piece-wise linear approximation to the cumulative distribution for the normalised emission coefficient $Q_{\rm{}abs}(\theta{})$, which can be directly inferred from our input tables. The random direction $(\theta{},\phi{})$ is then transformed from the reference frame of the local cell to the simulation frame to determine the emission direction for each emitted photon packet. Note that these random walk photon packets are only important for scattering into the line of sight, which has a negligible effect on our results; scattering at most contributes a fraction of $2\times{}10^{-6}$ to any pixel value in our synthetic images, and this only at the shortest wavelengths we consider (for $\lambda{}>100~\mu{}$m the scattering contribution is effectively zero for all pixels).

For peel-off photon packets, the outgoing photon direction is known, and the weight of the photon packet is determined from the anisotropic emission profile by first computing the angle $\theta{}$ between the given photon direction and the local magnetic field, and then using linear interpolation on the tabulated $Q_{\rm{}abs}(\theta{})$ profile.

For both cases, the approach above only gives us the Stokes $I$ intensity. The Stokes $Q$ intensities are set by linear interpolation on the tabulated values for $Q_{\rm{}abs,pol}(\theta{})$. Within SKIRT, polarisation vectors are stored normalised to the total intensity $I$, and keep track of the arbitrary reference direction that defines the polarisation frame. We always choose this reference direction so that it is perpendicular to both the local magnetic field direction and the propagation direction of the photon packet, so that Stokes $U$ is zero. The proper inter-mixing of $Q$ and $U$ within the polarisation frame of the synthetic observer is then treated upon scattering or detection of the photon packet as in \citet{2017PeestPolarizationMonteCarlo}. None of the processes we consider introduces circular polarisation, so that Stokes $V$ is always zero.

\subsection{Dust and source geometry}

Our radiative transfer simulations use the galaxy models from the Auriga simulations. This is a suite of cosmological zoom-simulations of Milky Way type galaxies that includes an extensive sub-grid model, including a treatment for magnetic fields. For more details, see \citet{2017GrandAurigaProjectProperties}. \citet{2021KapoorSyntheticHighresolutionUV} recently post-processed the set of Auriga galaxies with SKIRT using an approach similar to \citet{2016CampsFarinfraredDustProperties,2018CampsDataReleaseUV} and \citet{2017TrayfordOpticalColoursSpectral}, and generated synthetic fluxes and images from UV to submm wavelengths for each galaxy. Our extracted dust and source geometry differ in two aspects from those of \citet{2021KapoorSyntheticHighresolutionUV}: we also extract the magnetic field vectors, and we mainly use the 6 level 3 simulations that have a higher resolution than the 30 level 4 simulations presented in \citet{2021KapoorSyntheticHighresolutionUV}. In short, our source geometry consists of two components, based on the star particles in the Auriga simulations: old stellar sources (age $>10$~Myr) are assigned an SED template from the \citet{2003BruzualStellarPopulationSynthesis} library, while young stellar sources (age $<10$~Myr) are assigned an SED from the MAPPINGS III SED family \citep{2008GrovesModelingPanSpectralEnergy}. The parameters of these components are set to the same values as selected in \citet{2021KapoorSyntheticHighresolutionUV}. The dust geometry is determined from the gas content of the Auriga simulations using two different dust allocation recipes. The first recipe, \texttt{recSF8000}, assigns dust to all gas with densities above a threshold density, $\rho{}_{\rm{}thr}=0.13~{\rm{}cm}^{-3}$, or with temperatures below a threshold temperature, $T_{\rm{}thr}=8000$~K \citep{2016CampsFarinfraredDustProperties}. This corresponds to star-forming gas, which means the dust geometry is limited to dense gas. The second recipe, \texttt{recT12}, assigns dust to all cells obeying \citep{2012TorreyMovingmeshCosmologyProperties}
\begin{equation}
    \log_{10}\left(\frac{T}{{\rm{}K}}\right) < 6 + 0.25\log_{10}\left(\frac{\rho{}}{10^{10}h^2{\rm{}M}_\odot{} {\rm{}kpc}^{-3}}\right).
\end{equation}
This second recipe also assigns dust to more diffuse gas. For both recipes, dust is assigned according to the local density and metallicity, $Z$, as $\rho_{\rm{}dust} = f_{\rm{}dust}Z\rho{}$. Since the different recipes lead to different fractions of gas eligible to host dust, we choose different values of $f_{\rm{}dust}$: $f_{\rm{}dust}=0.225$ for recipe \texttt{recSF8000} and $f_{\rm{}dust}=0.14$ for recipe \texttt{recT12}. As in \citep{2021KapoorSyntheticHighresolutionUV}, we sample the dust distribution onto an adaptively refined octree grid with a maximum of 12 refinement levels and a maximum cell dust fraction of $10^{-6}$.

The spheroidal dust grains are assumed to align their short axis/axes with the magnetic field direction. The extracted magnetic field strengths all have values in the range $[0.001,800]~\mu{}$G. The highest magnetic field strengths are found in the galactic centre and our maximum value there is of the same order of magnitude as the $1~$mG field strength inferred in the Milky Way's central molecular zone \citep{2019MangilliGeometryMagneticField}. Since none of our extracted magnetic field vectors are very weak and since the densities in our models are low enough to allow full exposure to the interstellar radiation field that can generate radiative torques, we do not expect significant variations in alignment strength between cells, so that we assume a single alignment fraction $f_{S,A}$ for the entire dust mixture.

The dust alignment model introduced above only takes into account the diffuse galactic dust, and does not explicitly include cold dust hosted in GMCs and cold molecular clouds, which are not modelled on the Auriga resolution. Dust absorption and emission from these objects is however modelled implicitly through the use of the MAPPINGS III SEDs for star forming regions, which do contain a self-consistent treatment of dust, including self-absorption. These templates also model the radiation from O and B stars within the GMC that escapes into the interstellar radiation field, on top of the radiation from older stellar sources. Since the MAPPINGS III model does not include polarised dust emission, our model implicitly assumes that the emission from cold, dense dust is not polarised. The diffuse dust that we model explicitly is optically thin to its own radiation and therefore not significantly affected by self-absorption.

The full Auriga simulation suite contains 30 models of Milky Way like galaxies. These galaxies were modelled at a mass resolution of $\approx{}5\times{}10^4~{\rm{}M}_\odot{}$, referred to as \emph{level 4} resolution. 6 of these models were also run at 8 times better level 3 resolution (halos 6, 16, 21, 23, 24 and 27), and only 1 model was run at level 2 resolution (halo 6), which is another factor 8 higher. While none of these models is an accurate analogue of the Milky Way, some models have additional features (e.g. ongoing mergers) that make them less suited for a comparison with Planck data. Following \citet{2017PakmorMagneticFieldFormation, 2018PakmorFaradayRotationMaps}, we perform the bulk of our runs on the Auriga 6 model. The other Auriga models are then used to test the robustness of our results and to explore a wider variety of scenarios. Based on a resolution study presented in \appendixref{appendix:resolution}, we will limit ourselves to the six level 3 simulations.

\begin{table}
    \centering{}
    \caption{Bands used for the extrapolation simulations. For each band, we show the minimum, maximum and pivot wavelength. The B-BOP bands assume a simple top-hat transmission, for the other bands the actual filter transmission curves are used. The angular resolution, $\theta{}_a$, corresponds to the measured FWHM of the instrument point spread function. For the B-BOP bands, the quoted values correspond to the final preliminary estimates which differ slightly from the values quoted by \citet{2019AndreProbingColdMagnetised}. The resolutions quoted for ALMA correspond to the most compact antenna configuration from the Cycle 7 primer. This corresponds to the poorest resolution and the largest maximum resolved scales.}
    \label{tab:bands}
    \begin{tabular}{l c c c c}
    \hline{} \\
    Band name & $\lambda{}_{\rm{}min}$ ($\mu{}$m) & $\lambda{}_{\rm{}max}$ ($\mu{}$m) & $\lambda{}_p$ ($\mu{}$m) & $\theta{}_a$ ($''$) \\
    \hline{} \\
    B-BOP 70 & 52 & 88 & 75 & 7.6 \\
    B-BOP 200 & 135 & 225 & 217 & 21.7 \\
    B-BOP 350 & 280 & 420 & 340 & 37.9 \\
    \hline{} \\
    HAWC+ A & 45 & 66 & 53 & 4.85 \\
    HAWC+ C & 74 & 113 & 88 & 7.8 \\
    HAWC+ D & 119 & 210 & 155 & 13.6 \\
    HAWC+ E & 172 & 292 & 215 & 18.2 \\
    \hline{} \\
    SCUBA2 450 & 404 & 526 & 449 & 9.8 \\
    SCUBA2 850 & 750 & 994 & 853 & 14.6 \\
    \hline{} \\
    ALMA 10 & 320 & 380 & 349 & 0.43 \\
    ALMA 9 & 420 & 500 & 456 & 0.56 \\
    ALMA 8 & 600 & 780 & 689 & 0.88 \\
    ALMA 7 & 800 & 1089 & 937 & 1.23 \\
    \hline{}
    \end{tabular}
\end{table}

\begin{figure*}
\centering{}
\includegraphics[width=\textwidth]{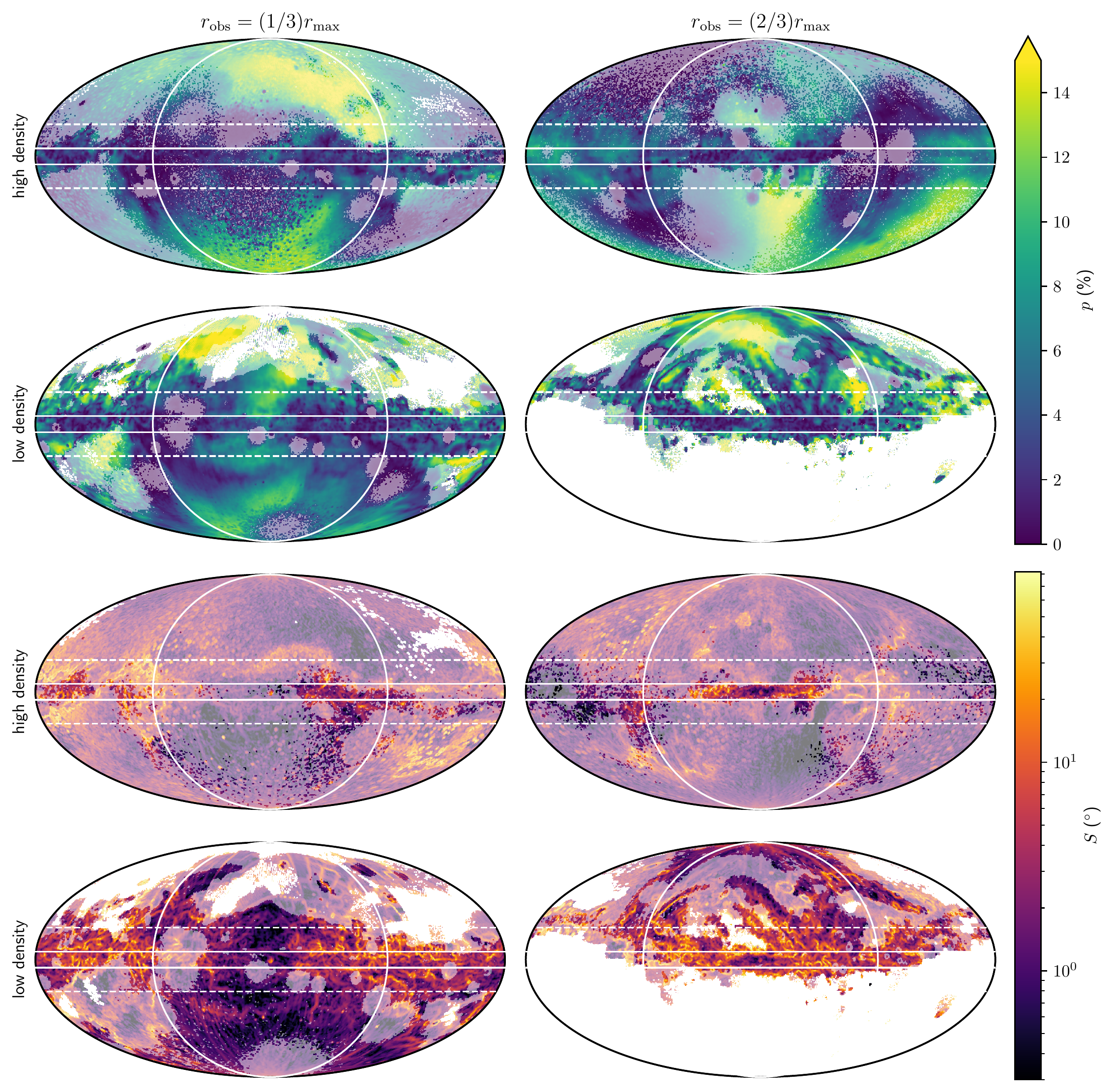}
\caption{Synthetic linear polarisation fraction (\emph{top rows}) and polarisation angle dispersion function (\emph{bottom rows}) across the entire sky for four observer positions within the galactic disc of the Auriga 6 model. The model shown here assumes a silicate-only dust mix with $f_{S,A}=0.6$ and a maximum intrinsic linear polarisation fraction of $\approx{}15~\%$. Dust was allocated using recipe \texttt{recSF8000}. The observer position is chosen within a ring with the radius indicated in the header of each column, and is chosen either as the position of the cell with the lowest or the highest density, as indicated in the label for each row. For our analysis, we have masked out the parts of the sky that have not recorded sufficient Monte Carlo photon packets to be statistically significant. Parts of the sky that are affected by this masking have been shaded out. The non-shaded parts of the map were computed after masking and hence do not necessarily match the shaded map at the boundary. The maps have the centre of the Auriga galaxy at their centre, and further have the galactic longitude $l$ increasing from right to left in the horizontal direction. The galactic latitude $b$ increases from bottom to top in the vertical direction. The white lines indicate the limits at $l=\pm{}90^\circ{}$ and $b=\pm{}5^\circ{},\pm{}20^\circ{}$ used when comparing different sky portions. Note that high values in the top panels have been saturated to a maximum intrinsic linear polarisation fraction of $15~\%$, as indicated by the triangle in the colour bar.}
\label{fig:halo6_allsky_SF8000}
\end{figure*}

\begin{figure*}
\centering{}
\includegraphics[width=\textwidth]{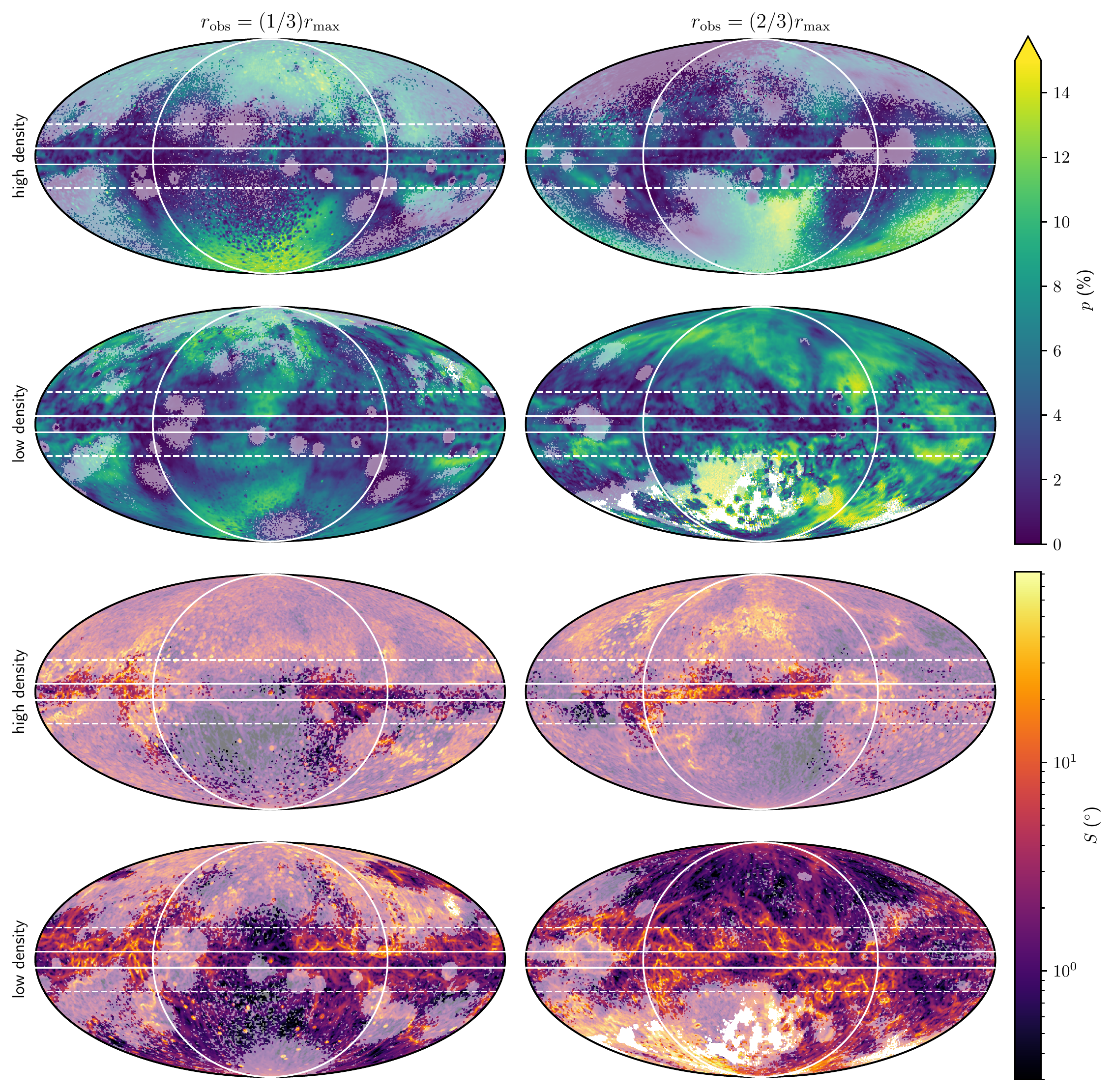}
\caption{Same as \figureref{fig:halo6_allsky_SF8000}, but for dust allocation recipe \texttt{recT12}.}
\label{fig:halo6_allsky_T12}
\end{figure*}

\begin{figure*}
\centering{}
\includegraphics[width=\textwidth]{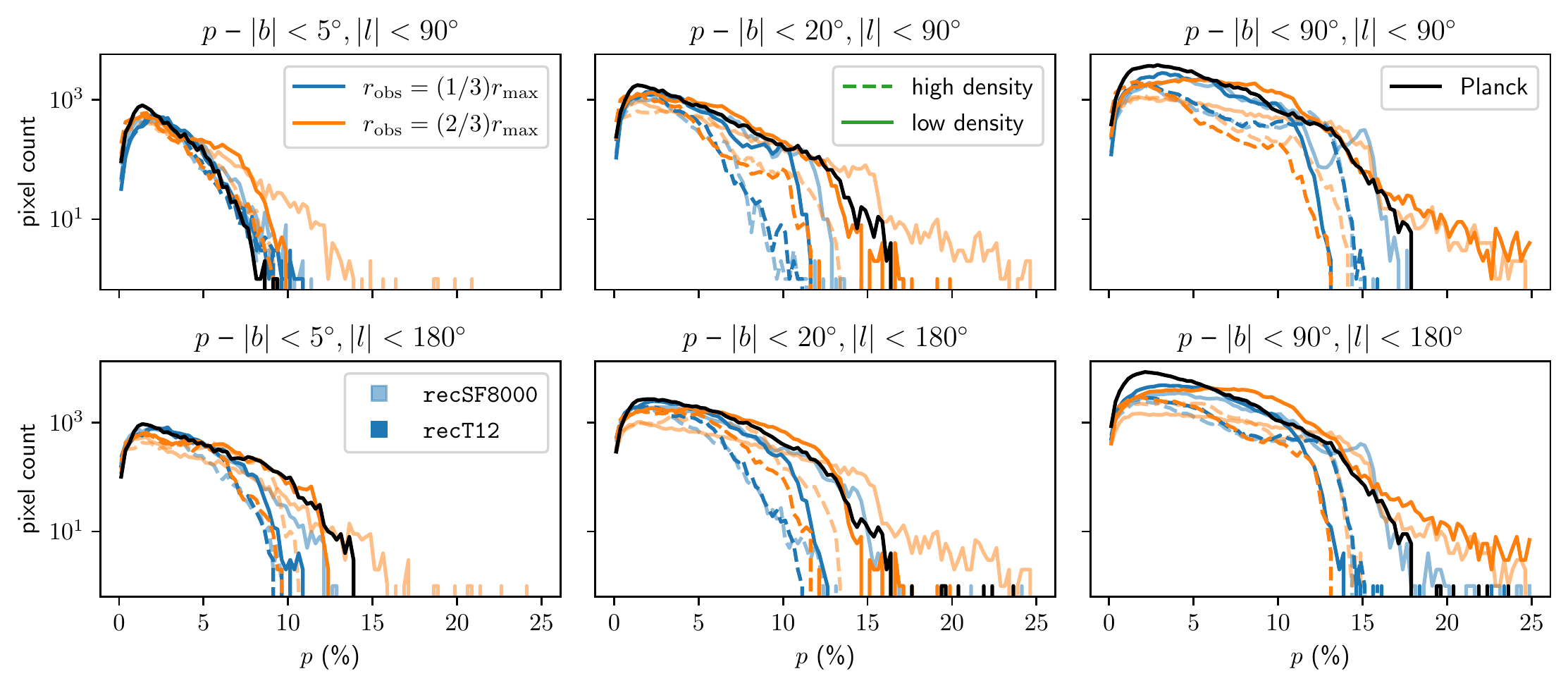}
\includegraphics[width=\textwidth]{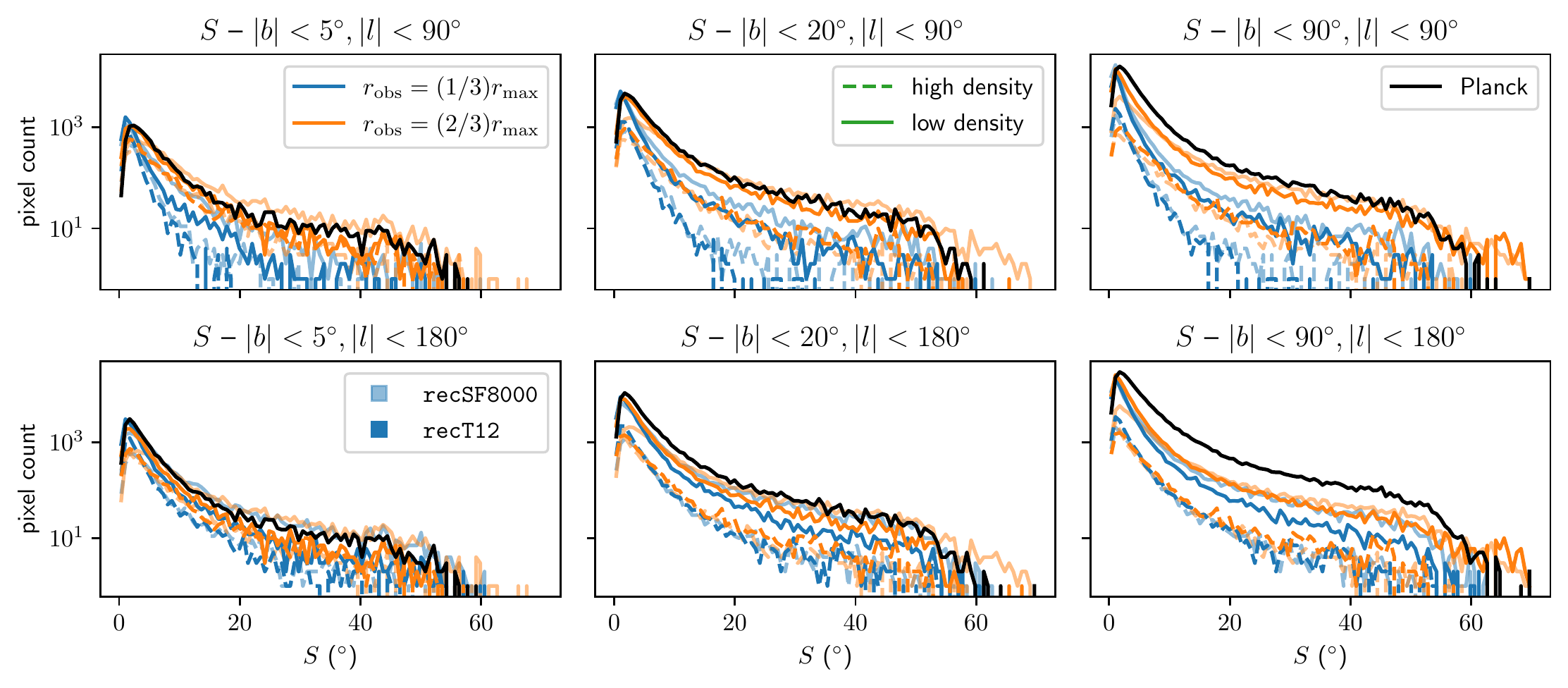}
\caption{Histograms of the linear polarisation fraction (\emph{top rows}) and polarisation angle dispersion function (\emph{bottom rows}) for six different cuts in galactic coordinates, as indicated above each panel. These histograms were computed for the masked images shown in \figureref{fig:halo6_allsky_SF8000} and \figureref{fig:halo6_allsky_T12} that use our reference dust model with alignment fraction $f_{S,A}=0.6$. The different colours correspond to different observer positions (columns in \multifigref{fig:halo6_allsky_SF8000}{fig:halo6_allsky_T12}), while the different line styles correspond to the different observer densities (rows in \multifigref{fig:halo6_allsky_SF8000}{fig:halo6_allsky_T12}). The opacity of the lines is determined by the dust allocation recipe, as indicated in the legend. The black solid line corresponds to histograms computed from the publicly available Planck HFI 353~GHz map, as presented in \citet{2020PlanckCollaborationPlanck2018Results}, using the same cuts in galactic coordinates. All histograms show the actual pixel count; no normalisation is performed.}
\label{fig:compare_positions}
\end{figure*}

\begin{figure*}
    \centering{}
    \includegraphics[width=\textwidth{}]{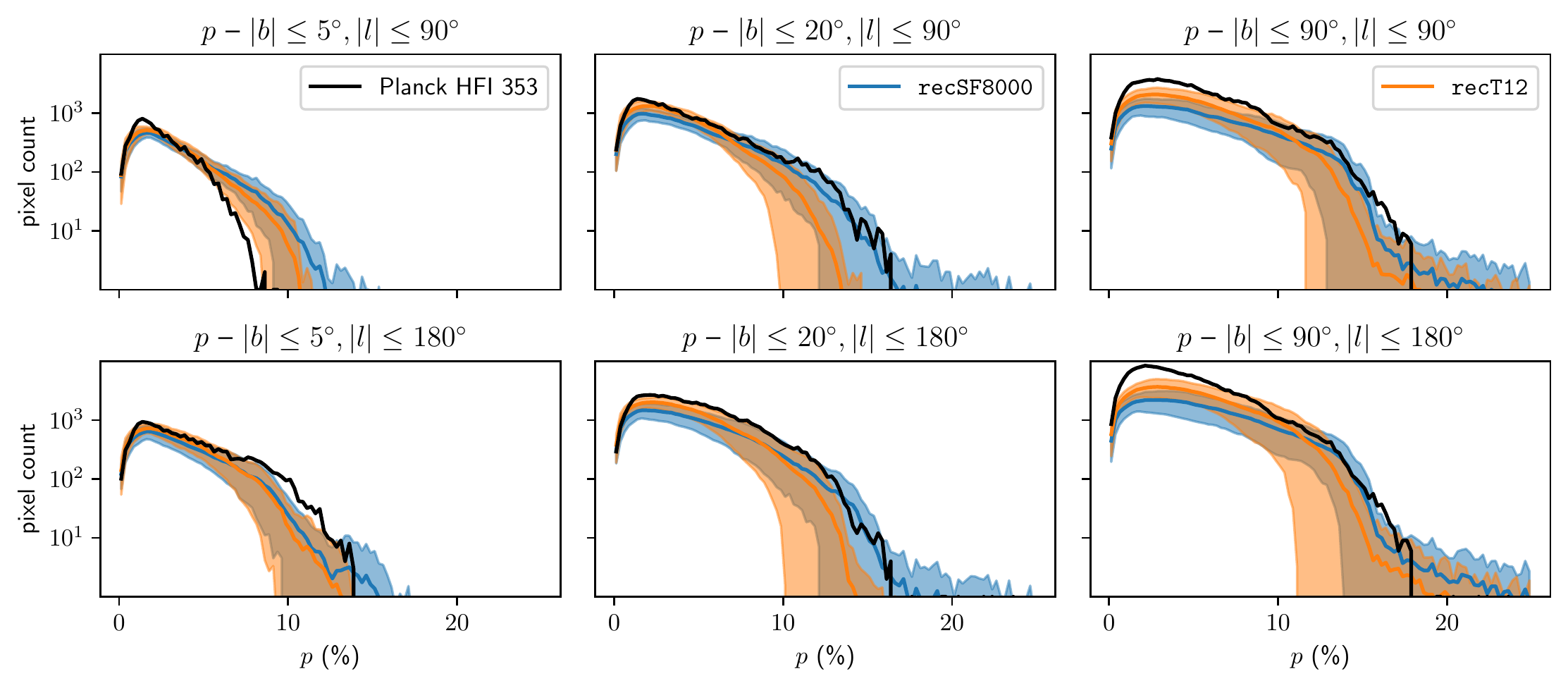}
    \includegraphics[width=\textwidth{}]{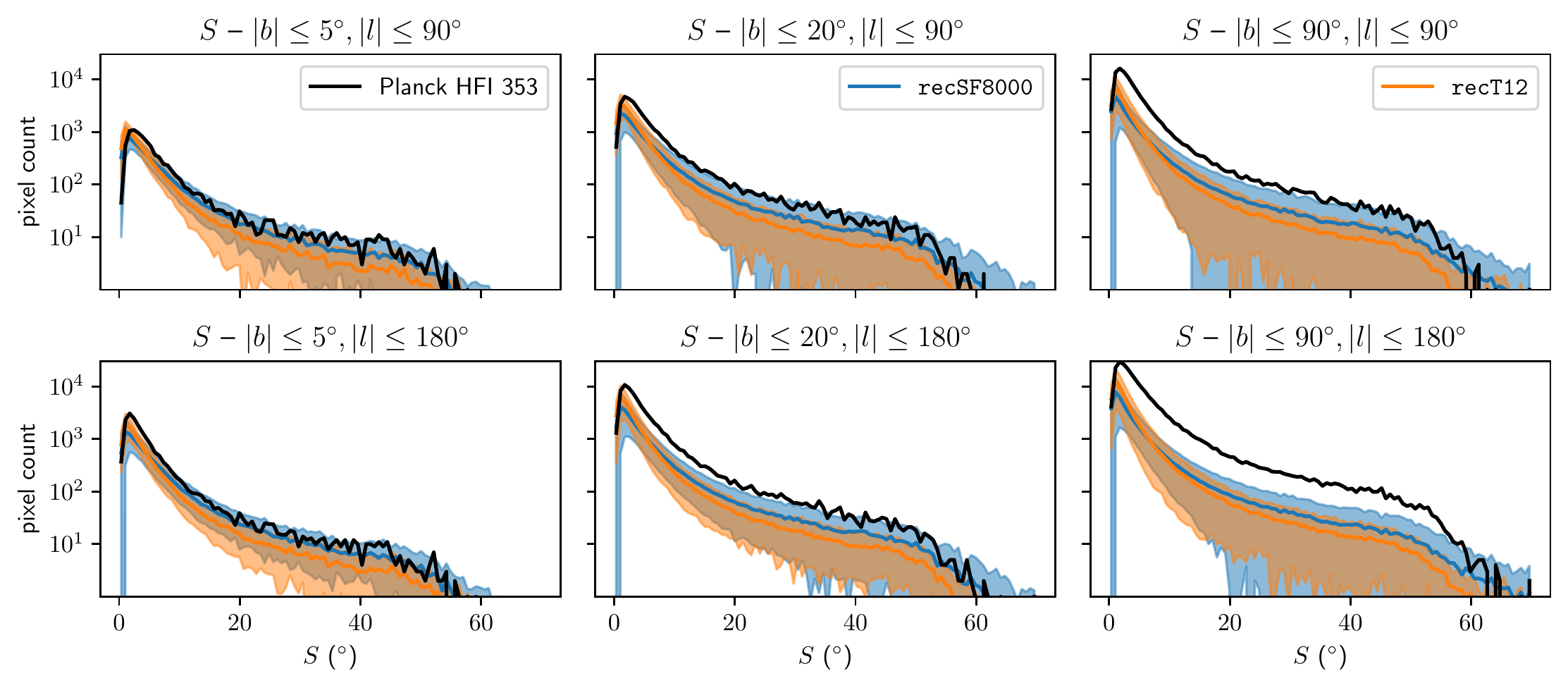}
    \caption{Same as \figureref{fig:compare_positions}, but now showing all models, averaged out over all 6 level 3 halos and all 4 observer positions and colour coded by the dust allocation recipe. The solid lines indicate the average in each histogram bin, while the shaded region quantifies the standard deviation in each bin.}
    \label{fig:compare_dust_allocation_models}
\end{figure*}

\begin{figure}
    \centering{}
    \includegraphics[width=0.49\textwidth]{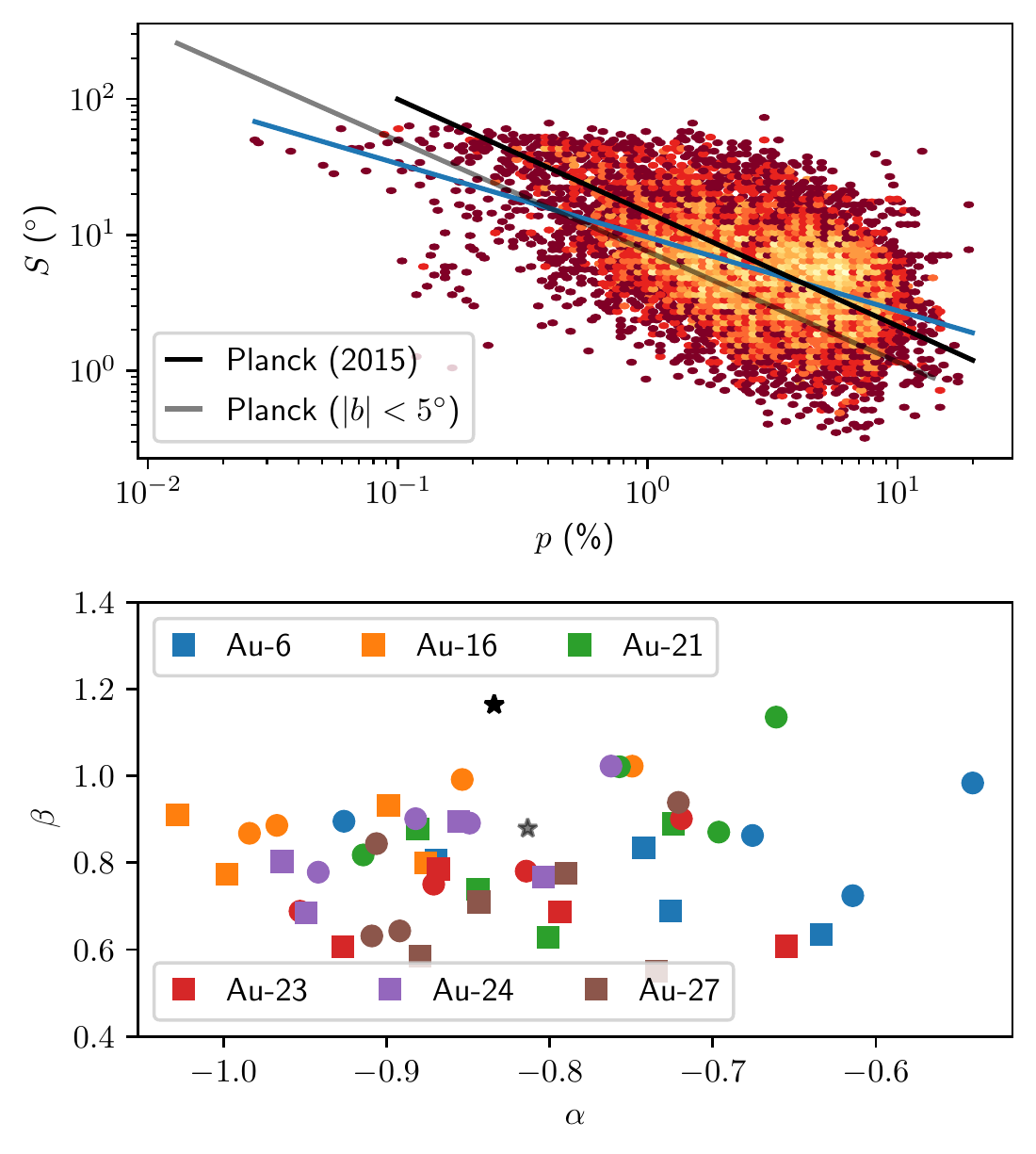}
    \caption{\emph{Top:} angular dispersion function as a function of linear polarisation fraction for all pixels with $|b|<5^\circ{}$ in the Auriga 6 model with dust allocation recipe \texttt{recSF8000} and an observer at the low density position in the outer annulus. Pixels have been binned to visualise the pixel density, with yellow indicating more pixels on a logarithmic scale. The blue line is a linear fit to the pixel values of the form $\log_{10}(S/^\circ{}) = \alpha{}\log_{10}(p/\%)+\beta{}$. The black line correspond to the \citet{2015PlanckCollaborationPlanckIntermediateResultsa} fit, while the grey line is a fit to the Planck data from \citet{2020PlanckCollaborationPlanck2018Results}. \emph{Bottom:} fit coefficients $\alpha{}$ and $\beta{}$ for the linear fits shown in the top panel, but now for all level~3 Auriga halos, as indicated in the legend. Discs indicate models that use dust allocation recipe \texttt{recSF8000}, while squares correspond to dust allocation recipe \texttt{recT12}. The stars correspond to the Planck data fits, with the same colours as in the top panel.}
    \label{fig:p_vs_S}
\end{figure}

\subsection{All-sky models}

We constrain the free parameter $f_{S,A}$ in our dust model against the polarised thermal emission for the Milky Way at 839~$\mu{}$m, as observed by Planck in the HFI 353~GHz band \citep{2015PlanckCollaborationPlanckIntermediateResultsa, 2020PlanckCollaborationPlanck2018Results}. A successfully calibrated dust model must satisfy some key Planck observations. First of all, Planck observed a maximum polarisation fraction $p_{\rm{}max}\approx{}22~\%$ and an anti-correlation between the column density along the line of sight and the polarisation fraction for that same line of sight, as predicted by \citet{2000HildebrandPrimerFarInfraredPolarimetry, 2008Falceta-GoncalvesStudiesRegularRandom}. This indicates that the intrinsic polarisation of the dust in the Milky Way has $p_{\rm{}max}>20~\%$, and that the Milky Way magnetic field is sufficiently tangled up to average out this signal along dense lines of sight. Second, Planck showed that the Milky Way magnetic field is relatively ordered on large scales, as measured from the polarisation angle dispersion function $S$, as proposed by \citet{2000HildebrandPrimerFarInfraredPolarimetry} and defined in \citet{2015PlanckCollaborationPlanckIntermediateResultsa}. $S$ also anti-correlates with the polarisation fraction, again indicating that the polarisation signal is lost due to line-of-sight averaging in regions with large field variations \citep{2018VaisalaSupernovaregulatedISMIV}.

Our comparison against the Planck Milky Way data serves a double purpose. If we are unable to reproduce the range of observed polarisation fractions and polarisation angle dispersion functions for any dust model, then this indicates that either the magnetic field structure in the Auriga simulations is not an accurate representation of the Milky Way magnetic field, or that our uniform dust model assumptions are too simplistic for our purposes. In this case, these simulations cannot be used to make predictions for the polarisation signal in disc galaxies. If however we are able to constrain our dust model, then this means we can use our zeroth order approximated dust model and the geometry of the Auriga simulations to extrapolate the Milky Way results to disc galaxies of arbitrary inclination, viewed externally and at shorter wavelengths.

Our all-sky runs require a choice of observer position within the galactic disc. Since none of the Auriga models is an exact analogue of our Milky Way, this choice of position is somewhat arbitrary. Rather than attempt to mimic the position of our Sun in the Milky Way, we select a range of observer positions using a simple but deterministic recipe. For each Auriga model, we determine a radius, $r_{\rm{}max}$, based on the size estimates from \citet{2021KapoorSyntheticHighresolutionUV}, defined as twice the radius at which the face-on stellar surface density within a 20~kpc layer around the midplane falls below a value of $2\times{}10^5~{\rm{}M}_\odot{}~{\rm{}kpc}^{-2}$. Based on this radius we select two annuli at a distance of respectively $r_{\rm{}max}/3$ and $2r_{\rm{}max}/3$ from the galactic centre and with thickness $0.2$~kpc. Within these annuli, we rank the cells based on their gas density, and place an observer at the position of the densest and the least dense cell. This mimics an observer located in a spiral arm and in an inter-arm region. This choice of positions will allow us to assess the impact of foreground extinction on our results.

For each observer position, we record the total intensity and Stokes Q and U intensity on a HEALPix\footnote{\url{http://healpix.sourceforge.net}} pixelisation of the celestial sky \citep{2005GorskiHEALPixFrameworkHighResolution, 2011GorskiHEALPixHierarchicalEqual}. To optimise the statistical significance of our results, we choose a relatively moderate HEALPix resolution of $0.5^\circ{}$, corresponding to a HEALPix side length parameter $N_{\rm{}s}=128$. This is significantly coarser than the full Planck resolution ($N_{\rm{}s}=2048$), but is still below the smoothing used by Planck to compute the polarisation angle dispersion function. We furthermore only model secondary emission in the Planck HFI 353~GHz band and restrict our secondary emission wavelength range to this band through means of a wavelength bias distribution \citep{2016BaesCompositeBiasingMonte}, which allows us to only generate photon packets that will be detected, with appropriately corrected statistical weights. Despite all these optimisations, our all-sky models suffer from a lack of cell resolution in the base Auriga simulations that makes it hard to obtain a good coverage at high altitudes.

Each all-sky simulation uses $5\times{}10^9$ photon packets to sample the primary source emission, and $5\times{}10^{10}$ photon packets for the thermal dust emission. The simulation self-consistently computes photon packet statistics for each image pixel, as described in \citet{2018CampsFailureMonteCarlo,2020CampsSKIRTRedesigningAdvanced}. Pixels that do not receive sufficient contributions to be statistically significant are masked out after smoothing the images with a Gaussian beam with FWHM $1^\circ{}$, consistent with Planck. For each observer position, we compute the linear polarisation fraction and the polarisation angle dispersion function. As in \citet{2015PlanckCollaborationPlanckIntermediateResultsa}, we use a lag of $0.5^\circ{}$ to compute the latter. Note that \citet{2020PlanckCollaborationPlanck2018Results} recommend using a $2^\circ{}$ smoothed polarisation map with a lag of $1^\circ{}$ to get a completely unbiased result, although their resulting polarisation angle dispersion functions show little statistical difference compared with the values computed using our adopted parameters. Since the publicly available GNILC maps used by \citet{2020PlanckCollaborationPlanck2018Results} are smoothed to $1^\circ{}$ resolution, we choose to use this base resolution instead of the additionally smoothed map.

\subsection{Extrapolation models}

Since this work was originally part of the preparations for the now cancelled ESA/JAXA mission SPICA \citep{2018RoelfsemaSPICAALargeCryogenic, 2020ClementsExplainESALastminute}, we extrapolate our successful dust models to wavelength ranges and resolutions compatible with the design of SPICA's B-BOP instrument \citep{2019AndreProbingColdMagnetised}, which consisted of 3 broad band filters centred on 70, 200 and 350~$\mu{}$m. These results can serve as guidance for future FIR missions \citep[e.g.][]{2018BattersbyOriginsSpaceTelescope}; this is especially true for the B-BOP 350 band that covers a gap in the coverage of current FIR instruments. We also provide predictions for a number of currently operational FIR polarimeters: the HAWC+ polarimeter on SOFIA \citep{2018HarperHAWCFarInfraredCamera} and the POL-2/SCUBA2 polarimeter on the JCMT \citep{2013HollandSCUBA210000}. While SKIRT can also provide predictions for the BLASTPol \citep{2008PascaleBalloonborneLargeAperture} and PILOT \citep{2016BernardPILOTBalloonborneExperiment} balloon borne experiments, we do not consider these instruments here. We also model the FIR ALMA bands that can observe polarised dust emission \citep{2016NagaiALMAScienceVerification}, with the important caveat that the maximum resolved scale of these bands in any ALMA configuration is too small to resolve the large scales in nearby galaxies. ALMA observations hence require observing in mosaic mode and are better suited for more distant galaxies. A full overview of all the bands included in our extrapolation simulations, their wavelength ranges and their angular resolution is given in \tableref{tab:bands}.

Our extrapolation simulations use the same dust model and Monte Carlo parameters used for the all-sky runs, but now record the thermal dust emission onto a regular rectangular image, mimicking a distant instrument. We record all values for an instrument at a distance of $10$~Mpc and at a resolution of $1000\times{}1000$ pixels for a field of view of $100\times{}100$~kpc. This captures the full extent of all level 3 Auriga models after data extraction. The images for different instrumental resolutions are smoothed in post-processing. Our nominal pixel resolution is $2''$, so that according to Shannon's theorem, we can accurately sample a beam with a FWHM of at least $\approx{}5''$ (corresponding to HAWC+ A; the highest resolution broad band in our study). We record images at three different inclinations, $i=0^\circ{}$ (face-on), $i=45^\circ{}$ and $i=90^\circ{}$ (edge-on), for one arbitrary viewing angle in each case. An important exception to the method outlined above are the 4 synthetic ALMA bands: since these have an angular resolution that is well below the pixel resolution of our $1000\times{}1000$ image, we rescale these to mimic observations at a distance of $100$~Mpc (effectively resulting in a spatial resolution similar to that of the HAWC+ A-D bands). We neglect small redshift effects that affect observations at this distance.

\section{Results}
\label{sec:results}

\subsection{Synthetic Planck maps}

\figureref{fig:halo6_allsky_SF8000} and \figureref{fig:halo6_allsky_T12} show all-sky maps of the linear polarisation fraction and the polarisation angle dispersion function for all observer positions in Auriga 6, for our reference dust model consisting of silicates with an alignment fraction $f_{S,A}=0.6$, and our two different dust allocation recipes. At all positions, there is a clear polarisation signal with a maximum polarisation fraction of $\approx{}15~\%$, and a maximum polarisation angle dispersion function of $\approx{}60^\circ{}$. Although quantitatively very different, the maps for both positions are qualitatively similar to the equivalent Planck maps for the Milky Way. The linear polarisation fraction is relatively low in the central plane, and reaches its maximum values at higher latitudes. The polarisation angle dispersion function exhibits the same spaghetti-like features observed in the Planck data, and shows similar spatial correlations. The models that use dust allocation recipe \texttt{recT12} generally have a more extended dust distribution, which translates into a better sky coverage of statistically significant pixels and a lower linear polarisation fraction at high galactic latitudes.

The combination of statistical noise and map smoothing tends to skew the linear polarisation fraction in our models towards values that are higher than the intrinsic maximum of our assumed dust distribution in regions of low statistical significance, illustrating the need for appropriate masking. This positive bias on $p$ due to sampling noise is similar to the positive bias caused by observational noise. While various techniques exist to correct for the observational bias \citep{2015MontierPolarizationMeasurementAnalysis, 2015MontierPolarizationMeasurementAnalysisa, 2019PattleJCMTBISTROSurvey}, correcting for the bias in our simulations is less straightforward, as we cannot compute the covariance matrix for the sampling noise. Because of the low density of high latitude dust in our dust geometry, obtaining good statistics at high latitudes requires an impractically high number of Monte Carlo photon packets. For this reason, our synthetic images are generally limited to a relatively thin disc, especially in the maps of the polarisation angle dispersion function. To enable a thorough comparison with the Planck data that is not hindered by differences in sky coverage, we perform a restricted statistical analysis on six different angular cuts in galactic longitude and latitude: two discs with latitudes $|b|<5^\circ{}$ and $|b|<20^\circ{}$, and two half-discs with the same latitudes but an additional vertical cut for $|l|<90^\circ{}$. We also consider a full-sky and a half-sky map with no cuts in latitude. The latter horizontal cuts are useful for images at position $2r_{\rm{}max}/3$, for which only pixels near the galactic centre are significant.

\subsection{Histograms}

\figureref{fig:compare_positions} shows histograms for the maps presented in \figureref{fig:halo6_allsky_SF8000} and \figureref{fig:halo6_allsky_T12} where we compare the pixel statistics for the various cuts in galactic coordinates to the Planck data for the Milky Way in the same regions of the sky \citep{2020PlanckCollaborationPlanck2018Results}. Since both our synthetic data and the smoothed Planck data use a HEALPix grid with side length parameter $N_{\rm{}s}=128$, the pixel counts can be directly compared. These histograms show a good agreement between our synthetic all-sky maps and the Planck sky. As expected, our synthetic maps have a significantly lower sky coverage than the Planck maps, especially for observer positions close to the galactic centre and in high density regions. For all maps, the sky coverage is best in the central cut with $|b|<5^\circ{}$ and $|l|<90^\circ{}$ that covers the centre of the galaxy. Most of the scatter that is observed in the histograms can be attributed to those differences in sky coverage. One notable exception is the large scatter in the linear polarisation fraction for the central cut in the \texttt{recSF8000} model for the low density observer in the outer annulus, which can be attributed to contamination by masked pixels, as is evident from the fact that this map shows linear polarisation fractions above the intrinsic limit of the dust model. Similar issues affect the full and half sky maps shown in the rightmost column in \figureref{fig:compare_positions}. We will therefore restrict ourselves to the cuts at lower latitudes for the remainder of our analysis.

A comparison of the histograms for different observer densities reveals that the statistical properties for different observer annuli radii are consistent, with differences being attributed to lower statistical significance in the high density environment because of foreground contamination. There is some indication of a systematic variation of the linear polarisation fraction with observer annulus radius, with observers at the inner annulus radius having lower polarisation angle dispersion functions, indicating a more ordered magnetic field structure perpendicular to the line of sight.

Overall, the maps for observers at the outer annulus radius are more consistent with Planck. The Planck histograms show a clear systematic reduction of the linear polarisation fraction in the inner plane, evident when comparing the central low latitude histogram with the full low latitude disc, which is not reproduced in our synthetic maps. The synthetic linear polarisation fractions consistently peak at higher values than the equivalent Planck data, and there are indications that the polarisation angle dispersion functions are on average lower than for Planck.

When we look at the histograms for all 6 level 3 halos in \figureref{fig:compare_dust_allocation_models}, averaged out over all 4 observer positions and all halos, then we see that our results for halo 6 are confirmed by the results for the other level 3 halos. There is some agreement between our synthetic maps and Planck, but we fail to reproduce the shift in the thin disc. The agreement between the two different dust allocation recipes is better, although there are some clear trends: recipe \texttt{recT12} yields lower linear polarisation fractions and has less scatter at high polarisation fractions. Recipe \texttt{recSF8000} appears to have a higher angular dispersion function. Both trends can be explained by the more diffuse dust distribution for recipe \texttt{recT12}. A more diffuse dust distribution leads to more line-of-sight averaging of the polarisation signal, which leads to lower linear polarisation fractions. A less diffuse dust distribution leads to a lower sky coverage of the polarisation signal, which leads to more noise at high linear polarisation fractions and reduces the signal in the angular dispersion function.

\subsection{Anticorrelations}

\citet{2015PlanckCollaborationPlanckIntermediateResultsa} found an anticorrelation between the linear polarisation fraction and the angular dispersion function for the same pixel. This anticorrelation is a consequence of the 3D structure of the magnetic field, since higher values of the angular dispersion function indicate a less ordered magnetic field perpendicular to the line of sight, while a lower linear polarisation fraction indicates a less ordered magnetic field along the line of sight within the instrument beam \citep{2018VaisalaSupernovaregulatedISMIV, 2020PlanckCollaborationPlanck2018Results}. In \figureref{fig:p_vs_S}, we see a similar anticorrelation in the synthetic Auriga images, which we quantify with a linear fit to the pixel values in log-log space. To reduce the impact of noisy pixels on our fits, we have limited our analysis to the thin disc, $|b|<5^\circ{}$. As can be seen in the bottom panel, our fit coefficients span a considerable range in slopes, with higher values for dust allocation recipe \texttt{recSF8000} that can be attributed to more noise in the linear polarisation fraction. In general, the intercept $\beta{}$ is lower than in the Planck fits, which indicates that our angular dispersion function is on average too low compared with Planck. This means the magnetic field perpendicular to the line of sight is more ordered in the Auriga simulations than it is in the Milky Way.

\begin{figure}
    \centering{}
    \includegraphics[width=0.49\textwidth]{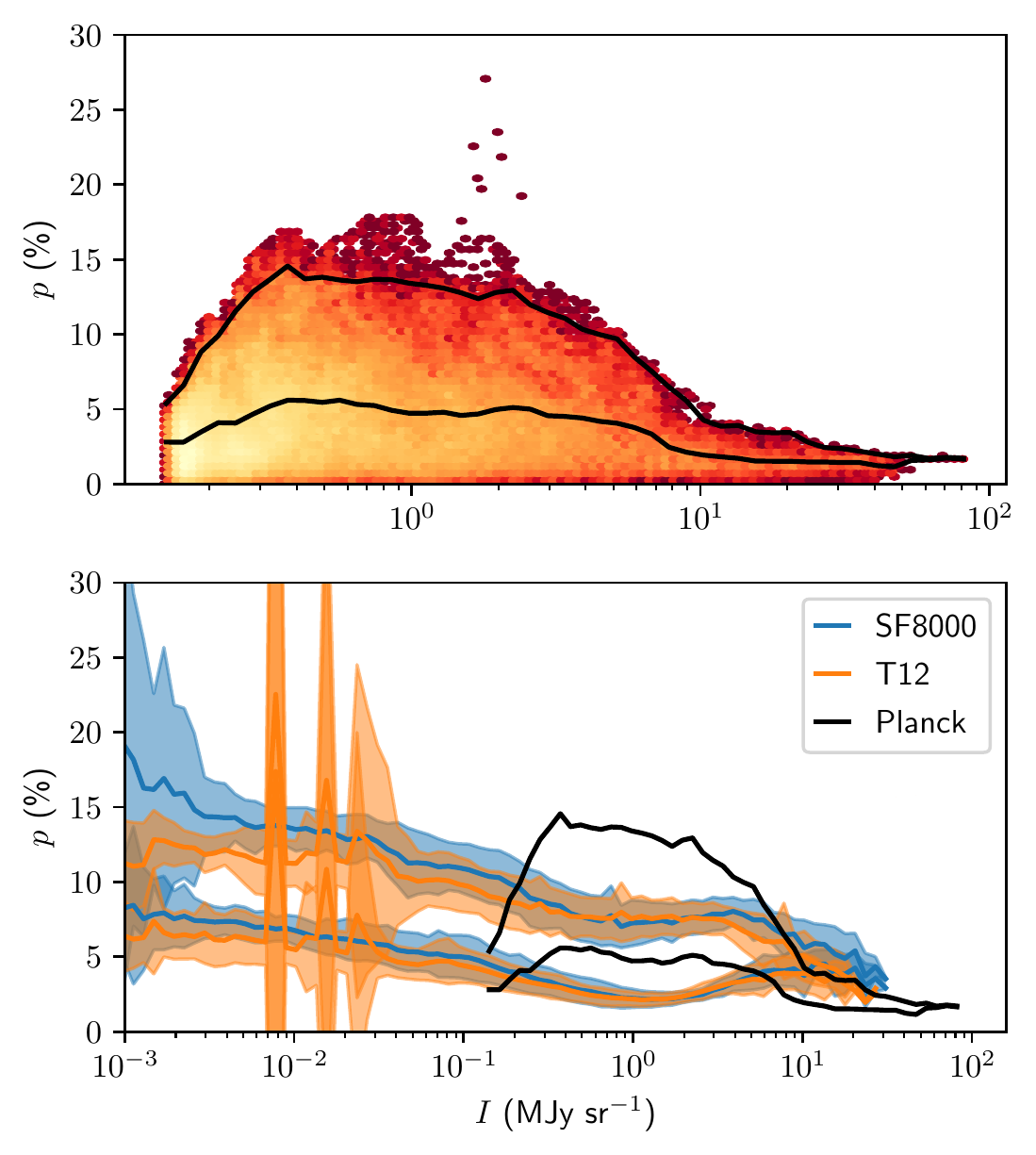}
    \caption{Linear polarisation fraction as a function of total intensity. \emph{Top:} Planck results for the full sky. Pixels have been binned to visualise the pixel density, with yellow indicating more pixels on a logarithmic scale. The black lines indicate the mean value for $p$ and the value for the $99~\%$ percentile of $p$ within 50 logarithmic bins in $I$. \emph{Bottom:} results from the 6 level 3 Auriga halos, again showing the mean and $99~\%$ percentile value of $p$ within 100 logarithmic bins in $I$. The solid lines indicate the average values of the mean and $99~\%$ percentile across all 6 halos and all 4 observer positions, while the shaded regions indicate the standard deviation across the 6 halos and 4 positions. The black lines are the same as in the top panel.}
    \label{fig:I_vs_p}
\end{figure}

\begin{figure}
    \centering{}
    \includegraphics[width=0.49\textwidth]{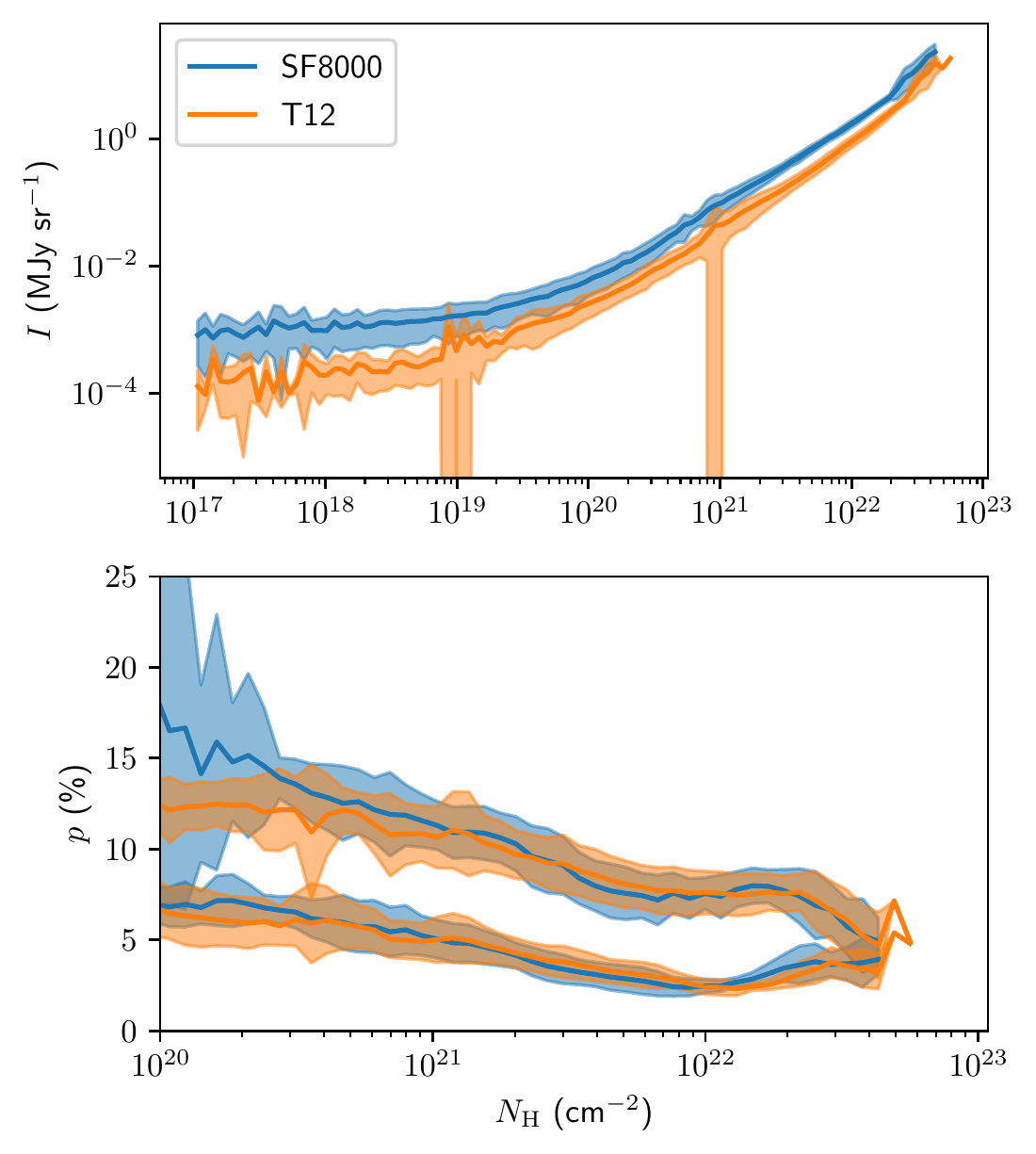}
    \caption{Total intensity (\emph{top}) and linear polarisation fraction (\emph{bottom}) as a function of hydrogen column density. All values have been binned within 100 logarithmic bins in $N_{\rm{}H}$. The solid lines always indicate the average of the binned quantity over all 6 halos and all 4 observer positions, while the shaded regions indicate the standard deviation across the halos and positions. Different colours correspond to different dust allocation recipes, as indicated in the legend. In the top panel, the binned quantity is the average total intensity. In the bottom panel, the binned quantities for the lower and upper lines correspond to respectively the mean linear polarisation fraction and the $99~\%$ percentile of the linear polarisation fraction in the bin.}
    \label{fig:NH_vs_p}
\end{figure}

\citet{2015PlanckCollaborationPlanckIntermediateResultsa} also found an anticorrelation between the optical depth along the line of sight and the maximum observed linear polarisation fraction for that optical depth, again indicating a loss of polarisation signal along dense lines of sight because of line-of-sight averaging. \citet{2019JonesSOFIAFarinfraredImaging} alternatively propose to examine the linear polarisation fraction in a pixel as a function of the total intensity in that pixel. This latter method has the advantage that it does not require a modified black-body fit to the intensity in different bands to obtain the optical depth, while still probing a similar dependency, since the total intensity $I$ approximately scales with the optical depth. The resulting graphs are shown in \figureref{fig:I_vs_p}. From the top panel, it is clear that the anticorrelation observed by Planck does indeed lead to a similar anticorrelation between $I$ and $p_{\rm{}max}$. A similar anticorrelation is found for the synthetic Auriga maps. However, while Planck found an average decrease of the linear polarisation fraction with increasing intensity, we find that the linear polarisation fraction instead increases with intensity for intermediate intensity values. Our linear polarisation fractions are systematically lower than the Planck data at the low end of the Planck intensities, and only reach their maximum values for very low intensities, well below the observed Planck range. It is worth noting that, although the synthetic maps in \multifigref{fig:halo6_allsky_SF8000}{fig:halo6_allsky_T12} reach intensities as low as $10^{-4}$~MJy~sr$^{-1}$, only $20~\%$ of the pixels has an intensity lower than $10^{-1}$~MJy~sr$^{-1}$, and only $5~\%$ of the pixels has values lower than $10^{-2}$~MJy~sr$^{-1}$. For high intensities, the observed drop in linear polarisation fraction is less strong than in the Planck data. The results for the different dust allocation recipes are very similar, with the most noticeable difference the fact that most \texttt{recSF8000} models have intensities well below the range shown in the figure, while \texttt{recT12} models generally have $I>10^{-4}~{\rm{}MJy~sr}^{-1}$. These differences between dust allocation recipes can be attributed to poor sky coverage and pixel noise in the \texttt{recSF8000} models.

Since we have the ISM density directly available from our model, we can also plot the linear polarisation fraction as a function of the actual column density along the line of sight, as is shown in \figureref{fig:NH_vs_p}. Note that these column densities are directly computed from the Auriga data cube and do not involve any radiative transfer. To facilitate the comparison with the smoothed intensity maps, we have smoothed the column density maps to the same $1^\circ{}$ angular resolution. It is clear from the top panel of \figureref{fig:NH_vs_p} that the total intensity correlates tightly with the column density along that line of sight. Also apparent is a systematic shift between the \texttt{recSF8000} and \texttt{recT12} dust allocation recipes, caused by a difference in normalisation that guarantees the same dust mass for the more extended \texttt{recT12} distribution. The linear polarisation fraction shows a similar dependence on the column density as observed by Planck \citep{2015PlanckCollaborationPlanckIntermediateResultsa}, with low values again being affected by pixel noise.

\subsection{Dust alignment fraction}

\begin{figure}
    \centering{}
    \includegraphics[width=0.49\textwidth{}]{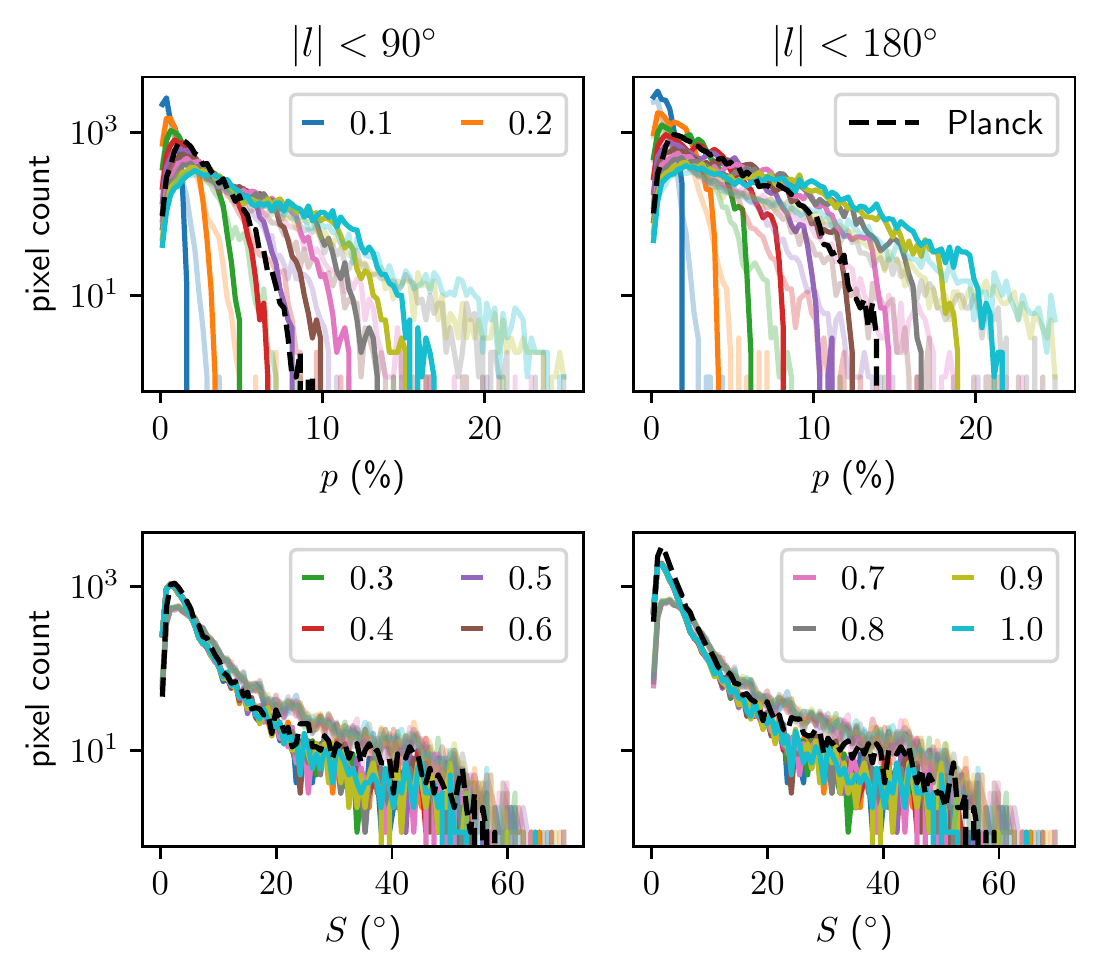}
    \caption{Histograms of the linear polarisation fraction (\emph{top}) and angular dispersion function (\emph{bottom}) for the Auriga 6 model and the observer at the low density position in the outer annulus. Only pixels in the thin disc with $|b|<5^\circ{}$ are shown. The different colours correspond to different values of the alignment fraction $f_{S,A}$, as indicated in the legend. The full lines are the models that use dust allocation recipe \texttt{recT12}, while the shaded lines are the models with dust allocation recipe \texttt{recSF8000}. The black dashed lines are the corresponding Planck histograms.}
    \label{fig:dustModel}
\end{figure}

\begin{figure*}
    \centering{}
    \includegraphics[width=\textwidth]{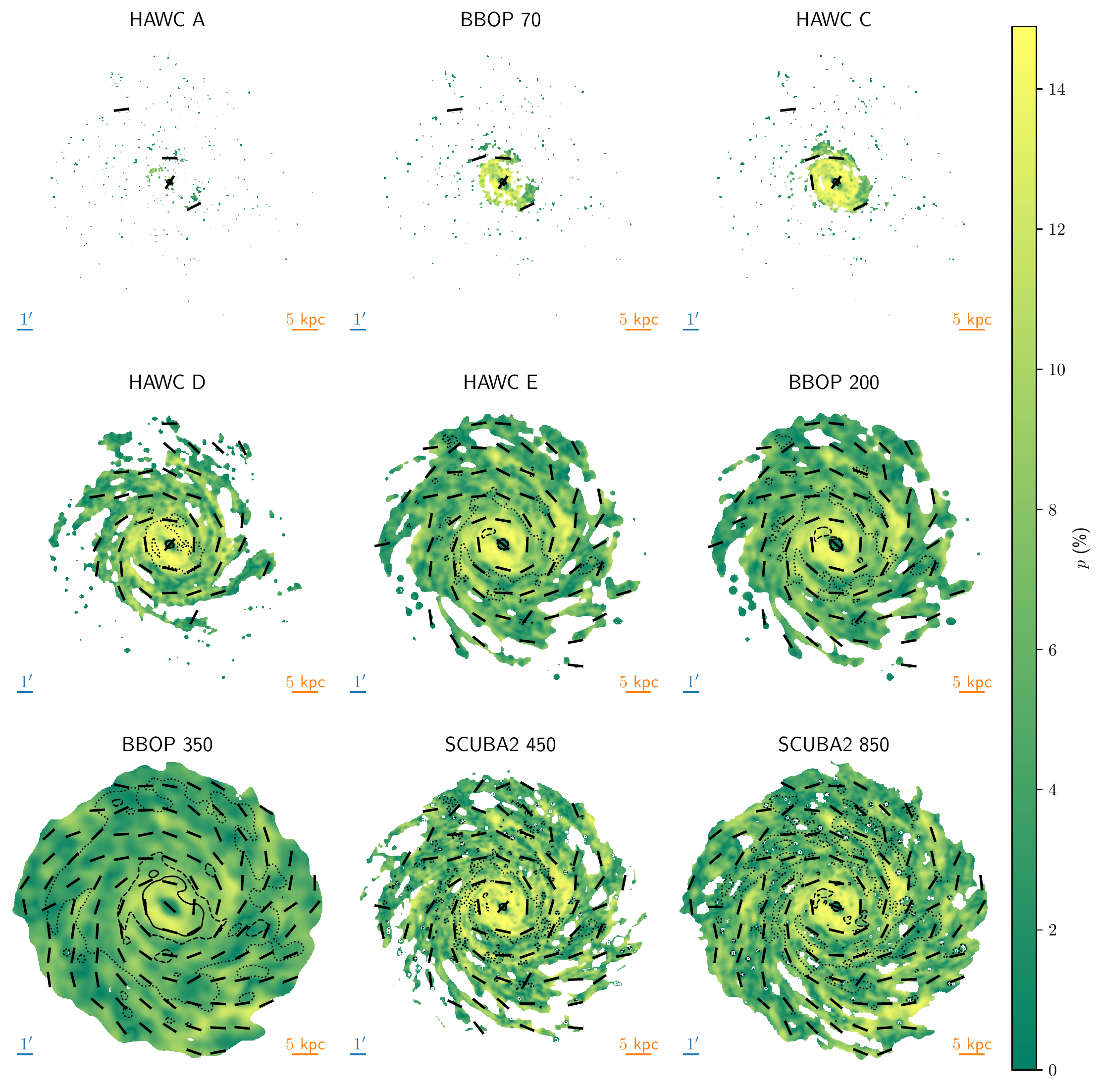}
    \caption{Linear polarisation fraction for the Auriga 6 reference model with $f_{S,A}=0.6$ and dust allocation recipe \texttt{recT12} for 9 synthetic broad bands and face-on inclination. The black dashes indicate the apparent magnetic field direction perpendicular to the line of sight, obtained by rotating the polarisation direction by $90^\circ{}$. The dotted, dashed and solid contours indicate regions with $I>[0.01,0.05,0.1] I_{\rm{}max}$ respectively. Pixels with $I<0.001 I_{\rm{}max}$ have been masked out. The angular and physical scale are indicated on the bottom of each panel; the latter assumes a distance of $10$~Mpc.}
    \label{fig:fig_halo6_face}
\end{figure*}
\begin{figure*}
    \centering{}
    \includegraphics[width=\textwidth]{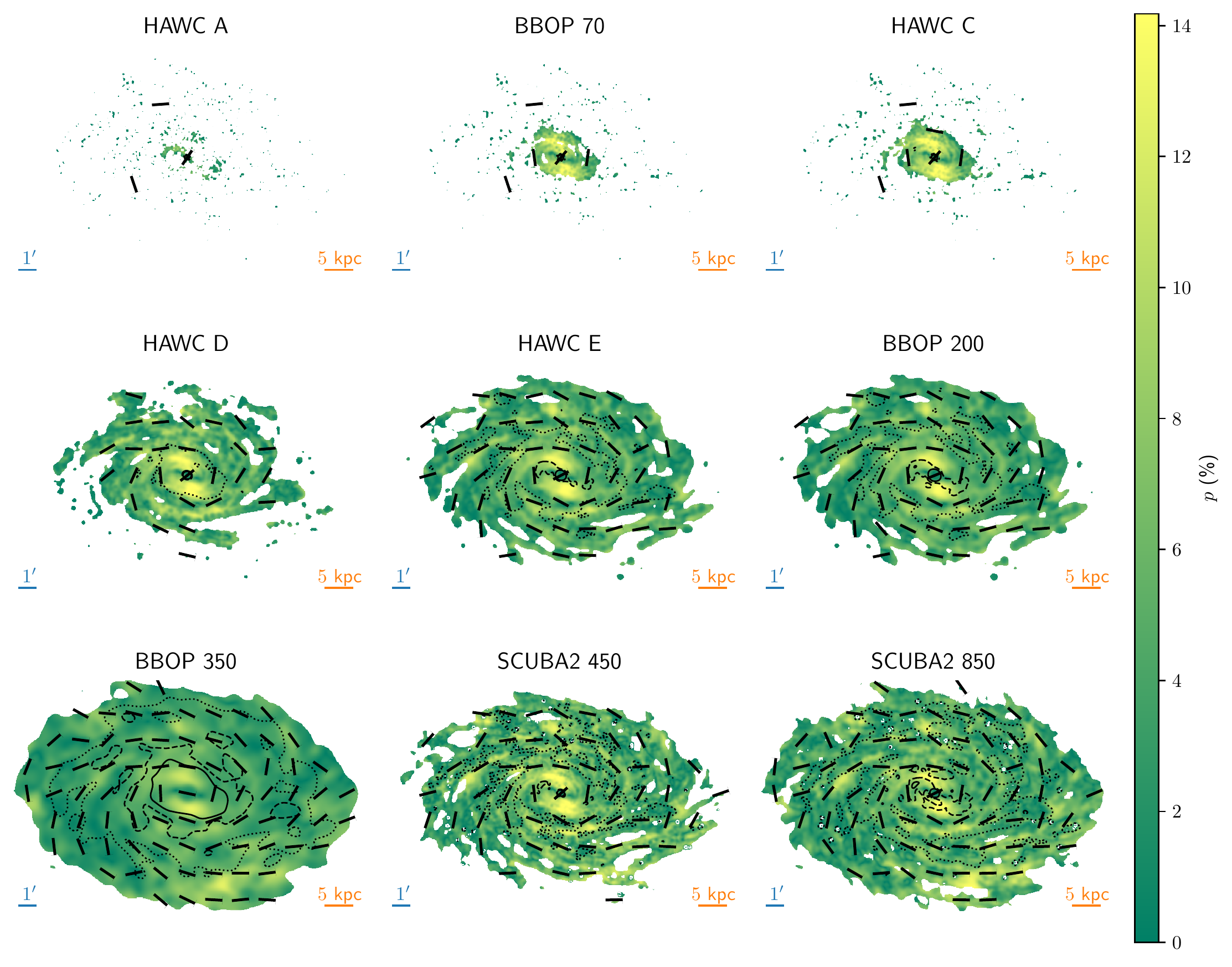}
    \caption{Same as \figureref{fig:fig_halo6_face}, but now showing an $i=45^\circ{}$ inclination.}
    \label{fig:fig_halo6_half}
\end{figure*}
\begin{figure*}
    \centering{}
    \includegraphics[width=\textwidth]{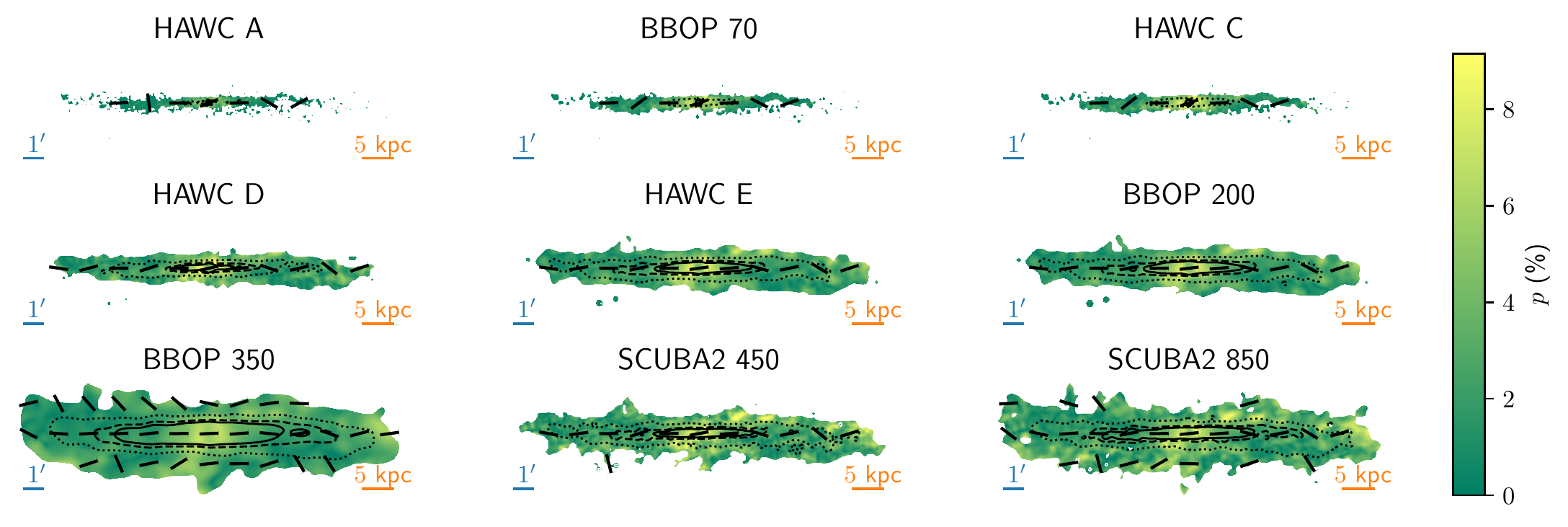}
    \caption{Same as \figureref{fig:fig_halo6_face}, but now showing an edge-on inclination.}
    \label{fig:fig_halo6_edge}
\end{figure*}

\begin{figure}
    \centering{}
    \includegraphics[width=0.49\textwidth{}]{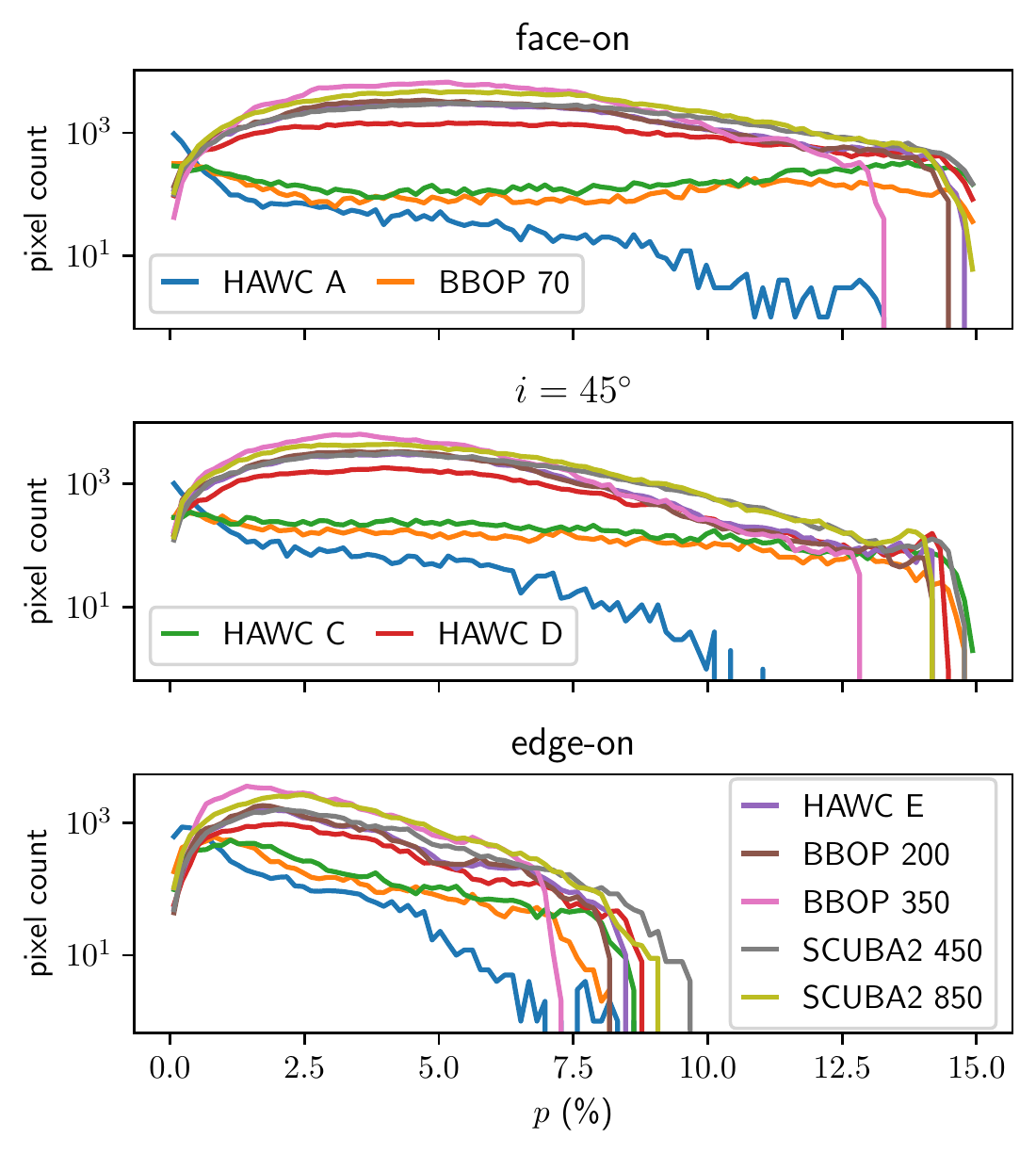}
    \caption{Histograms of the linear polarisation fraction as depicted in \multifigref{fig:fig_halo6_face}{fig:fig_halo6_edge}.}
    \label{fig:halo6_histograms}
\end{figure}

\begin{figure}
    \centering{}
    \includegraphics[width=0.49\textwidth{}]{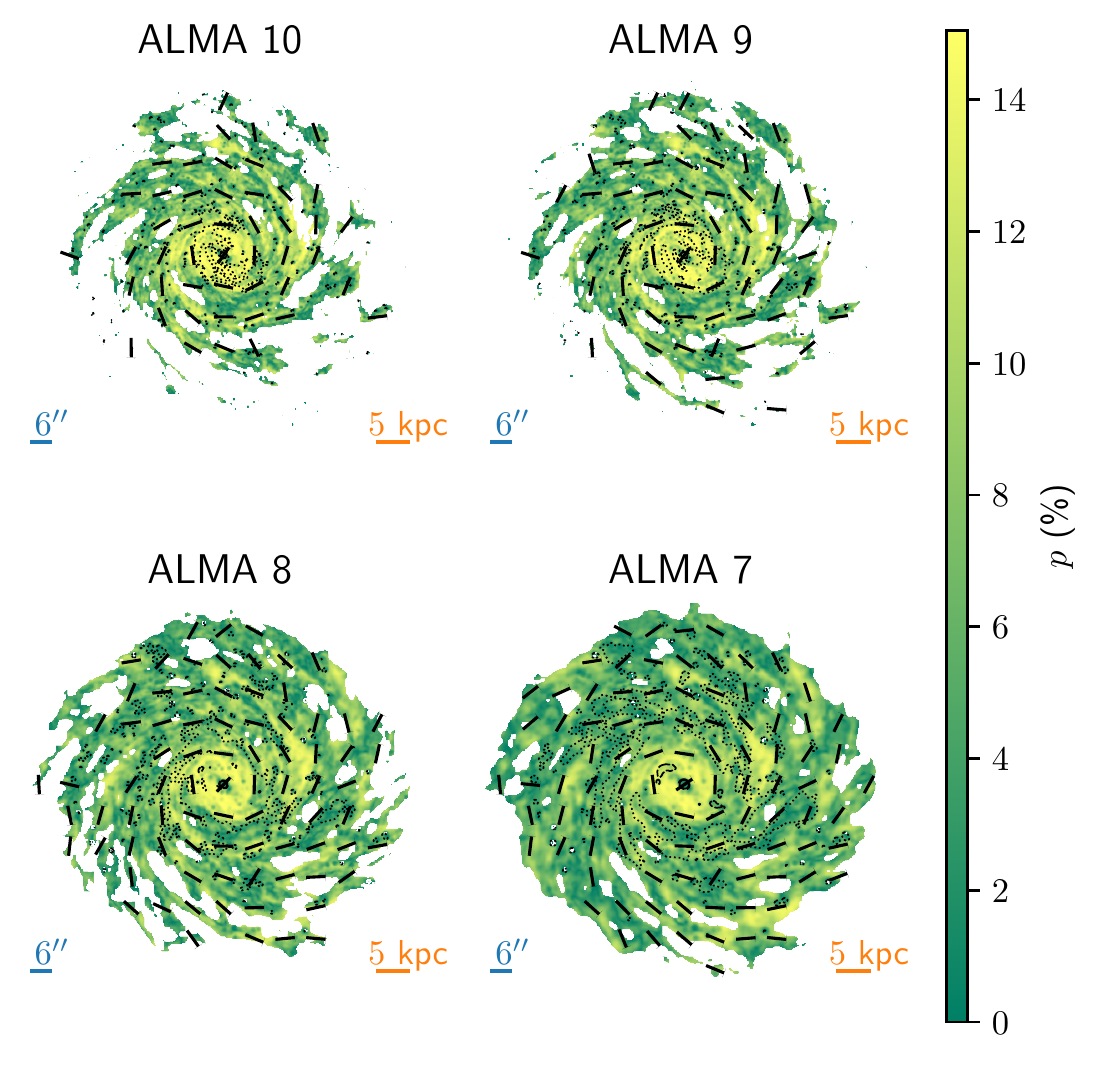}
    \caption{Same as \figureref{fig:fig_halo6_face}, but now showing the 4 synthetic ALMA broad bands for a face-on inclination, assuming an instrument distance of $100$~Mpc.}
    \label{fig:fig_halo6_ALMA}
\end{figure}

Most of the results shown in the previous subsection are a direct consequence of the geometry of the Auriga simulations. In this subsection, we investigate how our results depend on the one parameter which we allow to vary in our models: the alignment fraction $f_{S,A}$. As before, we only consider a silicate-only dust mixture. Given the general agreement between the different level 3 Auriga models, we will limit ourselves to the Auriga 6 model. We will focus our analysis on the thin disc, $|b|<5^\circ{}$ and look at cuts with $|l|<90^\circ{}$ and with no restrictions on the galactic longitude. Finally, we limit ourselves to the observer at the low density within the outer annulus, which has the best sky coverage in the thin disc.

\figureref{fig:dustModel} shows how the statistics of our synthetic maps change with $f_{S,A}$. As expected, a decrease in the alignment fraction leads to a decrease in the linear polarisation fraction as the intrinsic polarisation fraction of the dust model decreases. The angular dispersion function is robust against changes in the alignment fraction. This is also expected, since the angular dispersion function only traces the orientation of the polarisation signal, which is independent of its strength.

As before, we see that the Planck data are reasonably reproduced by our models. For the inner galactic disc, the linear polarisation fraction is best reproduced by our model with $f_{S,A}=0.5$, while the entire galactic disc better matches the reference $f_{S,A}=0.6$ model. Both dust allocation recipes yield similar results and favour the same alignment fraction when compared directly with Planck. As before, we notice a considerably larger amount of scatter in the \texttt{recSF8000} curves due to pixel noise.

Based on these results, we select the model with $f_{S,A}=0.6$ as our reference model, as it provides the best match for the global disc. We note that this choice might overestimate the central linear polarisation fraction by a few percent, a fact which we will need to bear in mind when looking at our predictions below.

\subsection{Nearby galaxy predictions}
\label{sec:extrapolations}

\multifigref{fig:fig_halo6_face}{fig:fig_halo6_edge} show the linear polarisation fraction and apparent magnetic field direction derived from the polarisation angle for the 9 synthetic bands (excluding the 4 ALMA bands) and the 3 different inclinations for an instrument at a distance of $10~$Mpc. All these models show Auriga 6, using our reference dust model with $f_{S,A}=0.6$ and dust allocation recipe \texttt{recT12}. The shortest wavelength bands are dominated by bright emission from small regions near the centre and show little magnetic field structure. At longer wavelengths a clear spiral magnetic field pattern is apparent. For the face on and $i=45^\circ{}$ view, the maximum linear polarisation fraction is highest in a ring surrounding the galactic centre. This maximum value is close to the intrinsic maximum of our dust model. The edge-on images have significantly lower linear polarisation fractions, with the largest values still found near the galactic centre. The apparent magnetic field direction aligns well with the galactic disc.

\figureref{fig:halo6_histograms} shows histograms of the linear polarisation fraction for the images shown in \multifigref{fig:fig_halo6_face}{fig:fig_halo6_edge}. Except for the bands at the shortest wavelengths, which have few significant pixels, most bands show a broad distribution, with maximum linear polarisation fractions up to the maximum of the intrinsic dust model. A notable exception is the B-BOP 350 broad band that has a lower maximum at $\approx{}13~\%$ and a more pronounced peak at lower linear polarisation fractions. This is caused by the relatively poor angular resolution of the B-BOP 350 band: we confirmed that the raw synthetic image before smoothing with the instrument beam does cover a wider range in linear polarisation fractions. Smoothing causes a mixing of pixels with different polarised intensities, leading to a net loss of polarisation signal. Similar effects are found if other synthetic bands are smoothed with larger beam sizes.

The 4 synthetic ALMA bands in our model have an angular resolution that is approximately a factor 10 higher than the other broad bands we consider. Combined with relatively small maximum resolved angular scales, this makes them impractical for use on the same targets as shown in \multifigref{fig:fig_halo6_face}{fig:fig_halo6_edge}. We can however use them to study more distant objects. \figureref{fig:fig_halo6_ALMA} shows the same model as in \figureref{fig:fig_halo6_face}, but now assuming an observer distance of $100$~Mpc, and ignoring small redshift effects that affect those distances. The same features that are observed in other broad bands at longer wavelengths are apparent, but now at a spatial resolution that is equivalent to that of the shortest wavelength broad bands in \figureref{fig:fig_halo6_face}.

\begin{figure}
    \centering{}
    \includegraphics[width=0.49\textwidth{}]{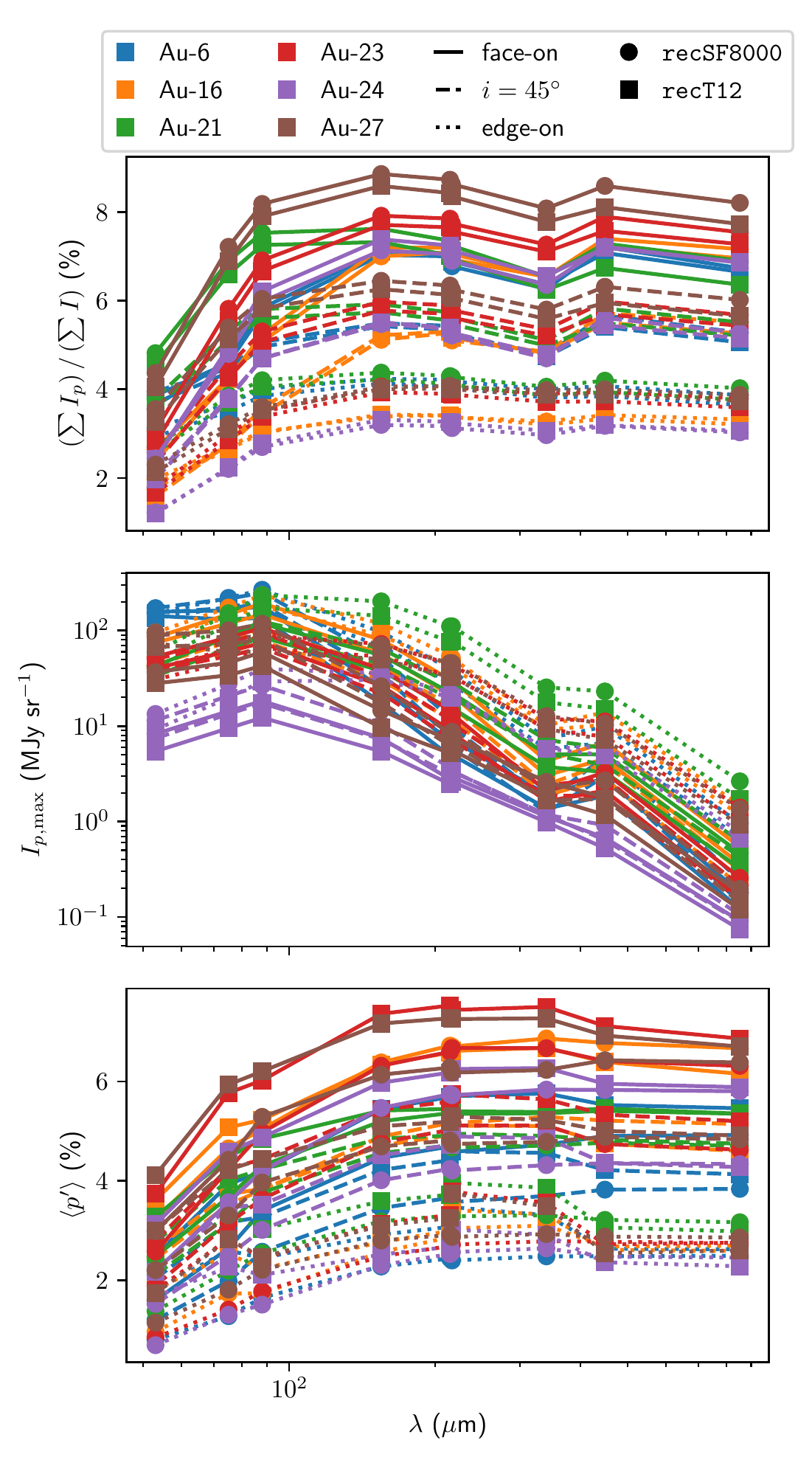}
    \caption{Intensity-averaged linear polarisation fraction (\emph{top}), maximum polarised intensity (\emph{middle}) and uniformly smoothed average linear polarisation fraction (\emph{bottom}) as a function of pivot wavelength for the 9 synthetic broad bands, 3 different inclinations and all 6 level 3 Auriga galaxies in our sample. The different colours correspond to different Auriga models, while the different lines indicate different inclinations and the different symbols indicate different dust allocation recipes. All models use our reference dust model with $f_{S,A}=0.6$. Note that the HAWC+ E and B-BOP 200 broad bands have very similar pivot wavelengths and are indistinguishable in this figure.}
    \label{fig:production_pstats}
\end{figure}

The top panel of \figureref{fig:production_pstats} shows the intensity-averaged linear polarisation fraction for all 6 level 3 Auriga galaxies, all 3 inclinations and both dust allocation recipes. The intensity-averaged linear polarisation fraction is computed as
\begin{equation}
    \langle{}p\rangle{} = \frac{\sum_i I_p}{\sum_i I},
\end{equation}
where the sum is over all pixels with $I>0.001I_{\rm{}max}$. This quantity has a clear dependency on wavelength, which is stronger than the wavelength dependence of the intrinsic dust model over the same wavelength range, consistent with \citet{2018GuilletDustModelsCompatible}. The intensity-averaged linear polarisation fraction is also a clear function of inclination and decreases for larger values of $i$. Finally, there is some dependence on the resolution of the observations, as clear from the lower values of $\langle{}p\rangle{}$ in the B-BOP 350 band as compared to the neighbouring B-BOP 200 and SCUBA2 450 bands that have higher angular resolution.

The middle panel of \figureref{fig:production_pstats} shows the maximum polarised intensity for the same models, defined as the maximum value of $I_p$ among all pixels with $I>0.001I_{\rm{}max}$. The general shape of this curve mimics the shape of the dust emission SED, but is also affected by the fact that shorter wavelength images have more concentrated emission. Since the polarised intensity is the quantity that is actually observed, this curve provides a good estimate of the observational accessibility of our model results. Our results for different Auriga galaxies show some spread depending on the overall luminosity (and size) of the different galaxies. For the same galaxy, the edge-on view has a higher maximum polarised intensity, and the maximum decreases for lower inclinations.

Resolution effects bias the wavelength dependence of the linear polarisation fraction. One potential way to correct for this bias is to smooth out all synthetic maps to the same angular resolution (we employ the B-BOP 350 resolution, $\theta{}_a = 37.9''$) and then compute the average linear polarisation fraction over the entire map, which we denote as $\langle{}p'\rangle{}$. This is shown in the bottom panel of \figureref{fig:production_pstats}. These uniformly smoothed average linear polarisation fractions reach peak linear polarisation fractions in a wavelength range $\approx{}[150,400]~\mu{}$m, and decrease for shorter and longer wavelengths. Since the linear polarisation fraction of the intrinsic dust model is constant over the full wavelength range covered by our synthetic broad bands, this wavelength dependence is entirely due to geometrical effects.

\begin{figure}
    \centering{}
    \includegraphics[width=0.49\textwidth]{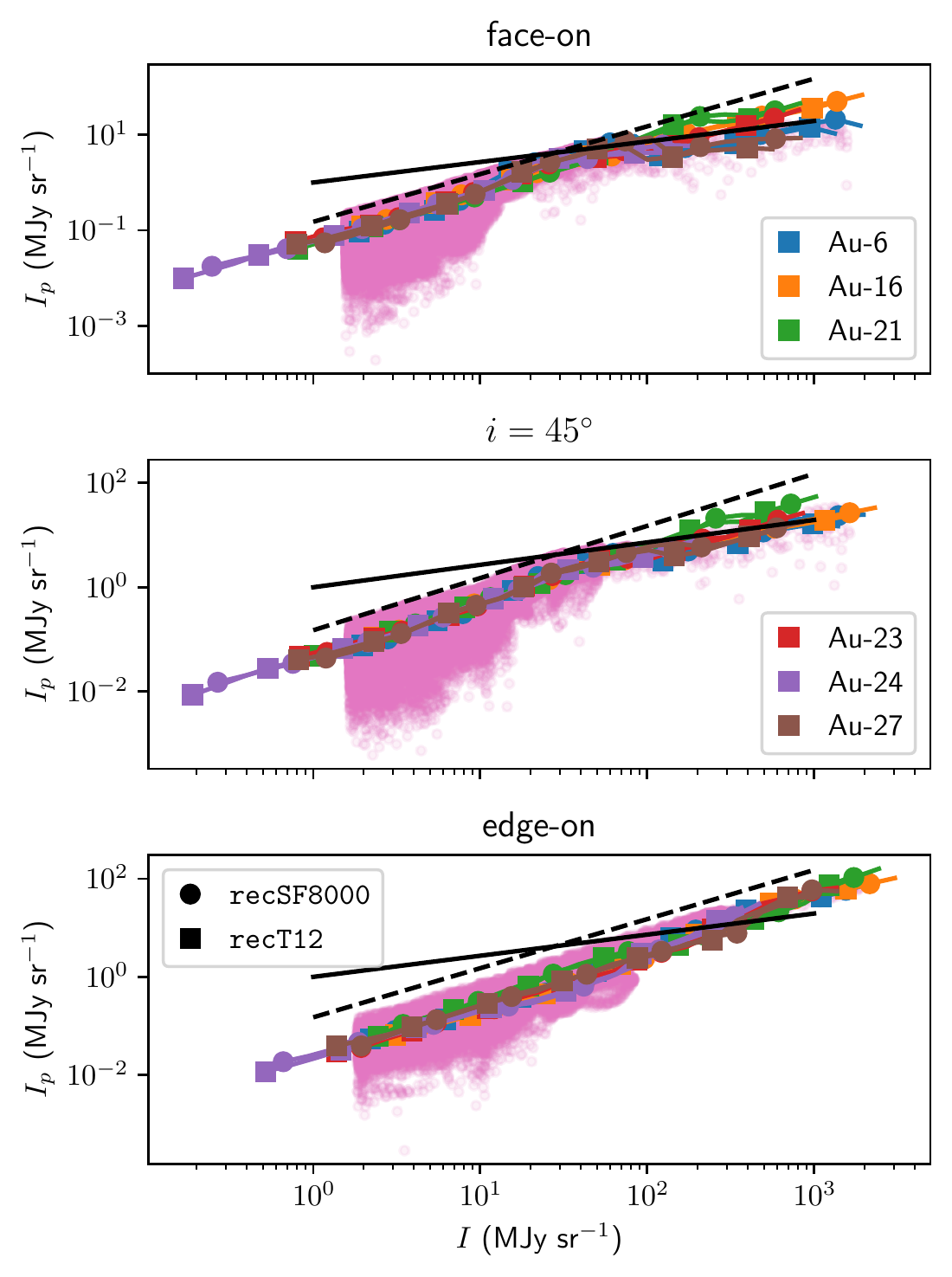}
    \caption{Polarised intensity as a function of total intensity for a face-on (\emph{top}), $i=45^\circ{}$ (\emph{middle}) and edge-on (\emph{bottom}) inclination, for all 6 level 3 Auriga galaxies using the two different dust allocation recipes, as observed in the synthetic HAWC+ D broad band. The transparent pink dots show individual pixel values for the Auriga 6 model with recipe \texttt{recT12}. The other coloured symbols correspond to the median polarised intensity in 20 logarithmic intensity bins, now for all models, as indicated in the legend. All curves have been limited to a minimum intensity of $I=0.001I_{\rm{}max}$. The different symbols correspond to the different dust allocation recipes. The solid black lines are the least squares fits to the M51 and NGC~891 data presented in \citet{2020JonesHAWCFarinfraredObservations}, which have practically the same slope. The dashed black line denotes the upper limit of the intrinsic dust model, $I_p=0.15I$.}
    \label{fig:production_I_vs_Ip}
\end{figure}

\figureref{fig:production_I_vs_Ip} shows the polarised intensity as a function of the total intensity, a relation which can be directly observed, see e.g. \citet{2020JonesHAWCFarinfraredObservations}. The scattered pink dots in the figure show the typical shape of this relation for one of our models. At low intensities, the polarised intensity shows large scatter. The upper limit at this end is set by the intrinsic dust model, as apparent in the top two panels. Note that the edge-on view in the bottom panel is also bounded by a straight line, but this line is well below the limit of the intrinsic dust model. This apparent shift in maximum linear polarisation fraction is expected when observing a transverse cylindrical magnetic field configuration on its edge, and a small similar shift is noticeable in the $i=45^\circ{}$ view. At the high intensity end, the polarised intensity shows less scatter. For the face-on and $i=45^\circ{}$ views, there is a clear kink at an intensity of $\approx{}10-100~{\rm{}MJy~sr}^{-1}$ where the polarised intensity no longer follows the intrinsic relationship.

To investigate this behaviour across our different models, we binned the polarised intensity in 20 logarithmic intensity bins and computed the median in each bin. These values are shown on top of the pink dots in \figureref{fig:production_I_vs_Ip}. While there is some scatter at the high intensity end, all models show a similar kink in the relation for the face-on and $i=45^\circ{}$ inclination. At the high intensity end, the relation agrees well with the least square fits to the M51 and NGC~891 data from \citep{2020JonesHAWCFarinfraredObservations}. Similar trends can be observed in other synthetic broad bands, albeit at different intensity levels.

\section{Discussion}
\label{sec:discussion}

\subsection{Comparison with Planck}

While our synthetic all-sky maps are not statistically equal to the Planck observations of the Milky Way, it is still remarkable how well we can reproduce the range and the overall shape of the histograms for the linear polarisation fraction and the polarisation angle dispersion function. The latter is especially remarkable, since this quantity is completely determined by the magnetic field structure from the Auriga simulations themselves, and was in no way part of our calibration process or the calibration process of the Auriga simulations. This confirms that the magnetic field structure in the Auriga simulations can be used as a realistic model for the large-scale magnetic field structure in a galaxy like our Milky Way, in line with the findings of \citet{2018PakmorFaradayRotationMaps} who constructed synthetic Faraday rotation maps for the Auriga galaxies.

Our inability to reproduce the increase in linear polarisation fraction for large galactic longitudes within the thin disc indicates a limitation of our model. Clearly, our synthetic model shows some level of isotropy when comparing the inner and the outer galactic disc which is not present in the Planck data. Additionally, there are clear indications that our dust emission is more diffuse than that observed by Planck, as evident from the generally lower intensities in our synthetic maps. This systematically lower intensity range is accompanied by a less outspoken decrease of the linear polarisation fraction with increasing total intensity. This indicates that we probe lines of sight with lower optical depths, but also that we observe less averaging of the polarisation signal along those lines of sight.

There are two possible explanations for this discrepancy. The first explanation is that the Auriga simulation lacks sufficient resolution to reproduce the dense filamentary structures in the centre of the galaxy that cause small-scale twisting of the magnetic field. The polarised emission from such a twisted magnetic field would, when averaged along the line of sight, lead to a reduction of the observed polarisation signal \citep{2000HildebrandPrimerFarInfraredPolarimetry,2020PlanckCollaborationPlanck2018Results}. Dense filaments would also boost the optical depth and the thermal dust emission along the line of sight. A lack of resolution could also explain why the model with $f_{S,A}=0.6$ provides the best fit to the observed Planck statistics. As evident from \figureref{fig:intrinsic_polarisation}, such a model has an intrinsic maximum linear polarisation fraction of only $15~\%$, well below the $p_{\rm{}max}\approx{}22~\%$ observed by Planck. If the line-of-sight resolution is underestimated over the entire sky, then this could lead to a systematic underestimation of the line-of-sight averaging which would be equivalent to using an intrinsic dust model with a lower alignment fraction. What argues against this explanation is that the Auriga resolution is in fact adaptive and therefore likely does not underestimate the line-of-sight resolution in a systematic way. Our limited convergence study in \appendixref{appendix:resolution} did also not indicate any significant shift in linear polarisation fraction for higher resolutions. While a lack of resolution could still conceal substructure in the densest unresolved clouds, which are modelled in a sub-grid fashion in the Auriga simulations themselves, it is therefore unlikely that resolution has a major impact on our large-scale results.

The second explanation is that our assumption of a constant dust model and alignment fraction for the entire galaxy is invalid. Models of grain alignment in dense discs \citep{2016ReisslRadiativeTransferPOLARIS} show that grain alignment becomes ineffective at high densities (although the densities involved in this work are much higher than considered here). Given the large complexity of grain alignment processes, it is not at all obvious that the alignment fraction would be constant, although a detailed analysis of the Planck data reveals no significant evidence for large-scale variations of grain alignment in the Milky Way \citep{2020PlanckCollaborationPlanck2018Results}. Even if silicate grains experience constant alignment on large galactic scales, there is still a possibility that gradients in the dust properties would alter the net polarisation signal in emission, which we also do not probe in our models. Detailed analyses of the FIR emission of nearby galaxies have revealed indications of radial variation in dust properties \citep[e.g.][]{2012SmithHerschelExploitationLocal,2019ClarkFirstMapsKd}, while observations of the Milky Way's central molecular zone show large variations in dust emissivity in the inner Galactic plane \citep{2019MangilliGeometryMagneticField} that could also be attributed to changes in the dust composition.

It is not possible to pinpoint the origin of this discrepancy within the limits of our current model. However, this discrepancy does introduce some uncertainty in our nearby galaxy predictions. If the observed radial trend in Planck is due to actual gradients in the polarised emission signal due to changes in the dust grain abundances or the grain alignment, then we would expect a clear radial trend in linear polarisation fraction in face-on galaxies. Since current FIR observations are limited to central parts of the galaxy, this would mean that our synthetic images overestimate the linear polarisation fraction in these observations. It is not clear if this would still be the case if the discrepancy is a resolution issue, since face-on observations probe significantly less dense material along the line of sight. In this case, our synthetic images could also under-predict the linear polarisation signal, since our synthetic all-sky maps somewhat under-predict the Planck data for the outer Milky Way.

Given the uncertainty about the resolution of our simulations, we cannot use our results to indicate a preferred dust allocation recipe. While in the context of our work, model \texttt{recT12} appears to perform better, the reasons for this better performance can ultimately be traced back to the extent of the dust distribution. A more extended dust distribution leads to more cells in our domain with a clear polarisation signal, which leads to more line-of-sight averaging and a better sky coverage. Given the relatively small differences between the two dust allocation recipes, one could assume that a higher resolution galaxy model could also yield more cells with a polarisation signal and achieve similar levels of line-of-sight averaging with model \texttt{recSF8000}.

\subsection{Nearby galaxy observations}

The main result presented in this work is highlighted in \figureref{fig:production_pstats}, which shows the strength of the expected polarisation signal for various bands and inclinations. The polarised intensity is strongest at relatively short wavelengths, but at these wavelengths the signal is dominated by compact regions and reveals little large-scale structure, as can also be seen in \multifigref{fig:fig_halo6_face}{fig:fig_halo6_half}. At these wavelengths, the linear polarisation fraction is also lower. At longer wavelengths, the emission is more extended, although the polarised intensity decreases by 2 orders of magnitude towards the SCUBA2 broad bands. The optimal combination of high polarised intensity and high linear polarisation fraction is obtained for pivot wavelengths $\lambda{}_p\in{}[100,300]~\mu{}$m, favouring the HAWC+ C, D and E bands and the B-BOP 200 band.

The B-BOP 350 band in \multifigref{fig:fig_halo6_face}{fig:fig_halo6_half} reveals a very similar magnetic field structure to the one probed in the B-BOP 200 band, but with a significantly lower polarised intensity, which is aggravated by the relatively poor resolution in this band, which smooths out the strongest polarisation signals. We therefore find no compelling arguments to use the B-BOP 350 bands in its final configuration in future FIR instruments for this kind of study. Turning the argument around, we find that observations of polarised emission are only able to trace the full extent of the distribution of the linear polarisation fraction if the angular resolution is $\approx{}20''$ at $10~$Mpc, corresponding to a minimum spatial resolution of $\approx{}1~$kpc.

ALMA has an angular resolution that is far superior to that of any of the other broad bands we considered in this work. As a result, it is in principle possible to use ALMA to trace the magnetic field structure in much more distant galaxies with the same spatial resolution as e.g. HAWC+ D observations of nearby galaxies. In practice, only ALMA band 7 has been used for polarisation observations to date. For nearby galaxies, ALMA observations are necessarily limited to small regions, making it harder to combine ALMA results with results obtained with other instruments.

When comparing \figureref{fig:fig_halo6_face} with \figureref{fig:fig_halo6_half}, we find that the magnetic field structure can still be traced in galaxies that are not inclined face-on, albeit with an on average lower linear polarisation fraction. Combined with the fact that the polarised intensity is higher for higher inclinations, this means that we expect to still obtain a good signal for nearby galaxies with reasonable inclinations.

\citet{2000HildebrandPrimerFarInfraredPolarimetry} presented histograms of expected linear polarisation strengths in astronomical objects at three different wavelengths, but noted that these results were obtained for envelopes of giant molecular clouds. In this regime, the linear polarisation fraction tends to decrease with wavelength, with values of only up to $5~\%$ at a wavelength of $350~\mu{}$m. The linear polarisation fractions in our synthetic images show a different trend: the HAWC+ A broad band with a pivot wavelength of $55~\mu{}$m has the lowest polarisation signal, strongly peaked at low linear polarisation fractions. Histograms for wavelengths $\lambda{}>150~\mu{}$m are more normally distributed with an average linear polarisation fraction of $\approx{}5~\%$ and maximum values beyond $10~\%$. Intermediate bands (B-BOP 70 and HAWC+ C) show the same broad distribution up to values beyond $10~\%$, but peak at significantly higher values, owing to a bias towards compact regions with high linear polarisation fractions. This discrepancy with the \citet{2000HildebrandPrimerFarInfraredPolarimetry} results can be attributed to selection effects, since we found that the highest linear polarisation fractions are recovered at low intensities which are harder to observe. But it can also be attributed to differences in the polarisation signal arising in galaxies, as also indicated by the relatively high values of the intensity-weighted average linear polarisation fraction we find.

Our results show reasonable agreement with HAWC+ observations of M51 and NGC~891 by \citet{2020JonesHAWCFarinfraredObservations}. We do recover the observed $I-I_p$ relation for face-on galaxies and additionally show a clear kink in the polarisation signal at high intensities which has not been probed by observations. \citet{2020JonesHAWCFarinfraredObservations} explain the high intensity relation as an imprint of line-of-sight averaging on the polarisation signal. High intensity dust emission originates in dense clumps with a complex magnetic field geometry, so that the polarisation signal is more likely to get averaged out along the line of sight \citep{2000HildebrandPrimerFarInfraredPolarimetry}. The fact that we do resolve a clear change in behaviour at a characteristic intensity (and dust density) which is consistent among our models, seems to be in line with this explanation. Note that no observations have probed the polarised intensity at low intensities in external galaxies yet, so that more observations are needed to confirm the existence of a clear kink in the $I-I_p$ relation.

Our maximum observed polarisation fractions of $\approx{}15~\%$ are significantly higher than the maximum polarisation fraction of $9~\%$ observed by \citet{2020JonesHAWCFarinfraredObservations}, which is in turn low compared to the maximum polarisation fraction of $22~\%$ observed by Planck. The discrepancy between HAWC+ and Planck observations does not necessarily indicate a tension, since Milky Way regions that exhibit high polarisation fractions in emission have relatively low total intensities, well below the sensitivity limit of HAWC+. The observations presented in \citet{2020JonesHAWCFarinfraredObservations} almost exclusively trace dense lines of sight with a clear imprint of line-of-sight averaging and are not incompatible with a maximum polarisation fraction of $22~\%$ at significantly lower intensities. Our models do show significantly higher polarisation fractions in regions with high emission intensities, indicating that our nearby galaxy predictions also benefit from a lack of line-of-sight averaging.

Our edge-on results show a clear tension with the NGC~891 results. While in this case the polarisation geometry agrees well, there is no evidence of line-of-sight averaging at high intensities in our edge-on images, while such a signal is indeed observed in NGC~891.

Our assumed intrinsic dust model shows very little wavelength dependence over the range covered by our synthetic broad bands ($[45, 1089]~\mu{}$m). Yet, the linear polarisation fraction (averaged in various ways; \figureref{fig:production_pstats}) still reveals a significant wavelength dependence over the various synthetic maps, even when resolution biases are corrected by smoothing all maps to the same angular resolution. This is mainly caused by the significant differences in spatial distribution of dust at different temperatures; warmer dust is more centrally concentrated and hence more affected by line-of-sight averaging. This agrees well with earlier models by \citet{2018GuilletDustModelsCompatible} that found a significant decrease in emitted polarisation fraction for wavelengths shorter than $200~\mu{}$m. Some of their models even showed a decrease in emitted polarisation fraction for wavelengths shorter than $300~\mu{}$m, which indicates a stronger impact of geometry than found in this work. Observations of nearby galaxies at different wavelengths are hence not a good tool to trace the wavelength dependence of the intrinsic polarisation of their interstellar dust, since there is significant contamination by geometrical effects.

\section{Conclusion}
\label{sec:conclusion}

In this work, we presented synthetic observations of polarised dust emission in nearby galaxies, based on the Auriga galaxy simulation models. We find that the linear polarisation signal of the Milky Way as observed by Planck at $839~\mu{}$m can be reasonably reproduced by a dust distribution consisting of pure silicate grains that are $60~\%$ aligned with the local magnetic field. We recover the observed distribution of linear polarisation fractions across the sky, although our results overestimate the linear polarisation fraction near the galactic centre by a few percent. The polarisation signal in our synthetic images gets weaker with increasing optical depth along the line of sight and correlates well with the magnetic field coherence perpendicular to the line of sight, in line with Planck observations.

When extrapolated to synthetic images of nearby galaxies, our models predict a clear imprint of the magnetic field structure at wavelengths above $\approx{}150~\mu{}$m, as accessible with the HAWC+ D and E bands and the SCUBA2 450 and 850 bands. We predict maximum linear polarisation fractions of up to $15~\%$ near the galactic centre, and we also show that these results can be obtained for galaxies with inclinations of at least up to $45^\circ{}$. The best window for observations of polarised emission is found for $\lambda{}\in{}[100,300]~\mu{}$m. To probe the full extent of the distribution function for the linear polarisation fraction, a minimum spatial resolution of $1~$kpc is required.

Our results predict a kink in the relation between the polarised intensity and the total intensity. The precise intensity value for which this kink occurs depends on the observed broad band. Below this intensity, the polarisation signal is no longer affected by line-of-sight averaging. To observe this kink, a dynamic range of at least two orders of magnitude in intensity needs to be observed.

The results presented in this work are limited by the spatial resolution of the Auriga models and the crude assumptions underlying our dust model. They do not account for spatial variations in grain alignment or grain properties, or the fact that dense regions in the Auriga simulations could be under-resolved. Better observations of polarised emission in nearby galaxies are required to assess the validity of our models before we can incorporate additional effects into our models.

\section*{Acknowledgements}

We thank Robert Grand for making the Auriga simulation data available to us. BV acknowledges funding from the Belgian Science Policy Office (BELSPO) through the PRODEX project ``SPICA-SKIRT: A far-infrared photometry and polarimetry simulation toolbox in preparation of the SPICA mission'' (C4000128500). This work used the DiRAC@Durham facility managed by the Institute for Computational Cosmology on behalf of the STFC DiRAC HPC Facility (www.dirac.ac.uk) as part of the allocation for project dp172. The equipment was funded by BEIS capital funding via STFC capital grants ST/P002293/1, ST/R002371/1 and ST/S002502/1, Durham University and STFC operations grant ST/R000832/1. DiRAC is part of the National e-Infrastructure. This paper made use of open source software, including NumPy \citep{2020HarrisArrayProgrammingNumPy}, SciPy \citep{2020VirtanenSciPyFundamentalAlgorithms} and Matplotlib \citep{2007HunterMatplotlib2DGraphics}. Some of the results in this paper have been derived using the healpy and HEALPix package \citep{2019ZoncaHealpyEqualArea}.

\bibliographystyle{aa}
\bibliography{main}

\begin{appendix}

\section{Resolution}
\label{appendix:resolution}

\begin{figure*}
    \centering{}
    \includegraphics[width=\textwidth{}]{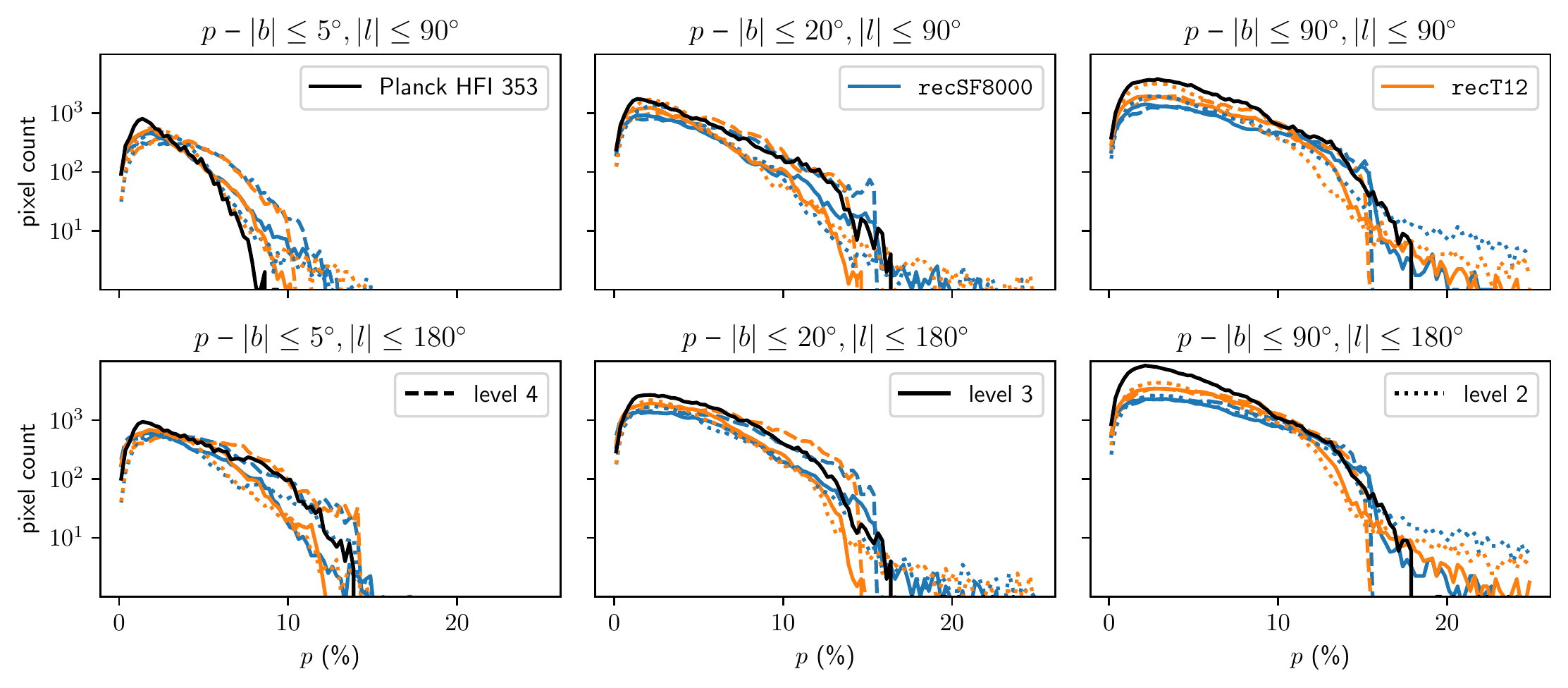}
    \includegraphics[width=\textwidth{}]{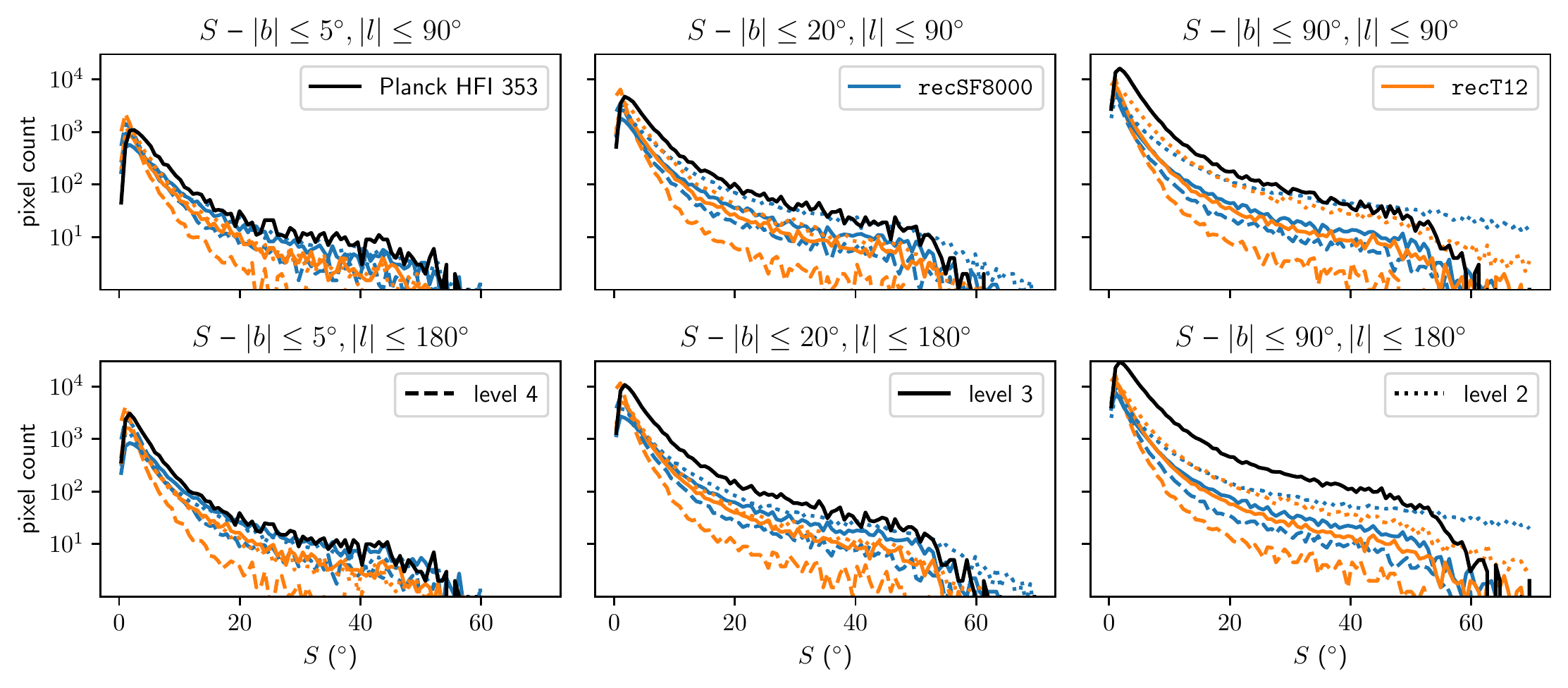}
    \caption{Same as \figureref{fig:compare_dust_allocation_models}, but now showing the average histograms for the Auriga 6 models at 3 different resolution levels, as indicated in the legend.}
    \label{fig:resolution_histograms}
\end{figure*}

\begin{figure}
    \centering{}
    \includegraphics[width=0.49\textwidth{}]{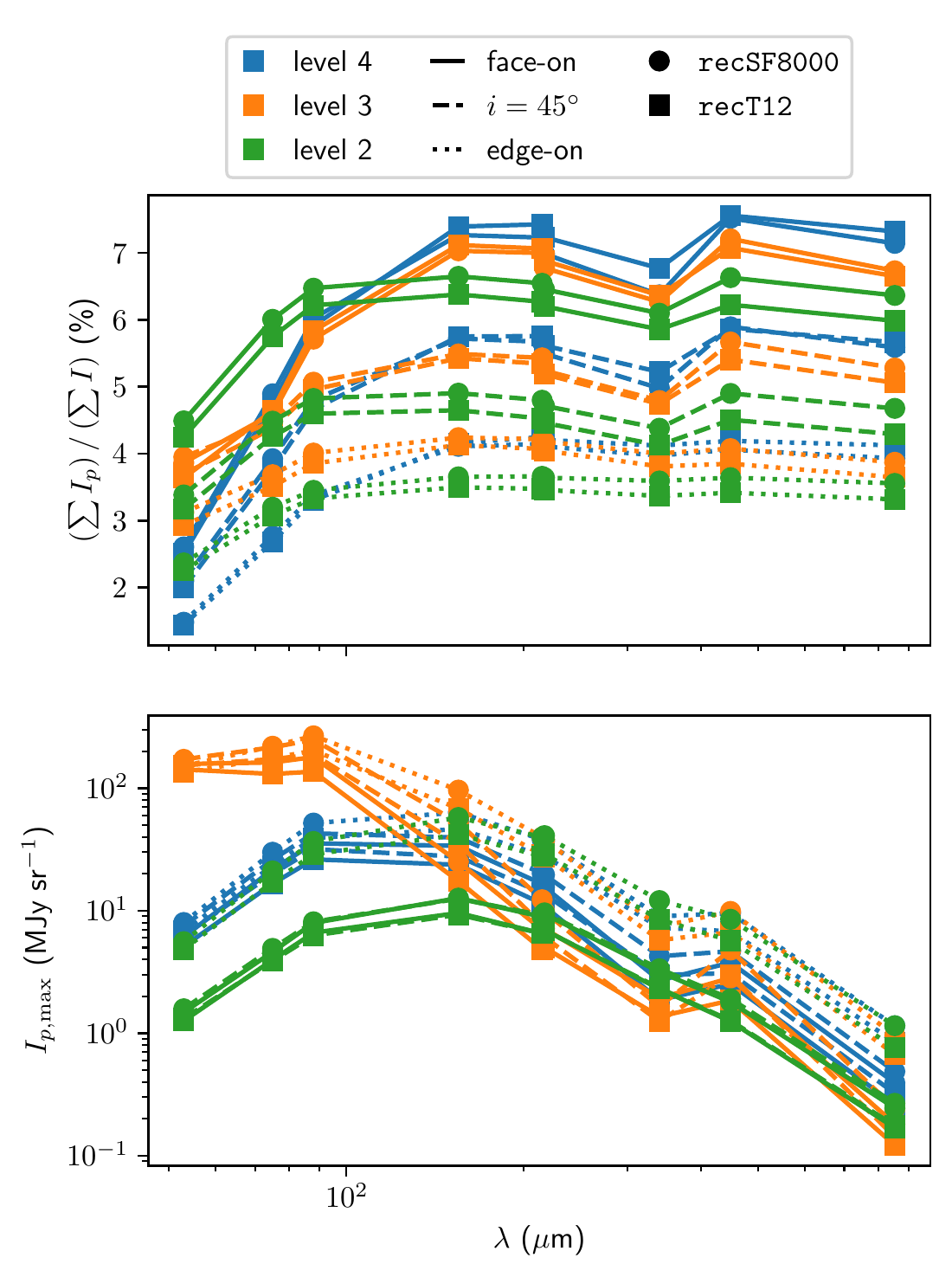}
    \caption{Same as the top panels of \figureref{fig:production_pstats}, but now showing the intensity-averaged linear polarisation fraction and maximum polarised intensity as a function of wavelength for the 3 different resolutions of the Auriga 6 galaxy.}
    \label{fig:resolution_pstats}
\end{figure}

\begin{figure}
    \centering{}
    \includegraphics[width=0.49\textwidth{}]{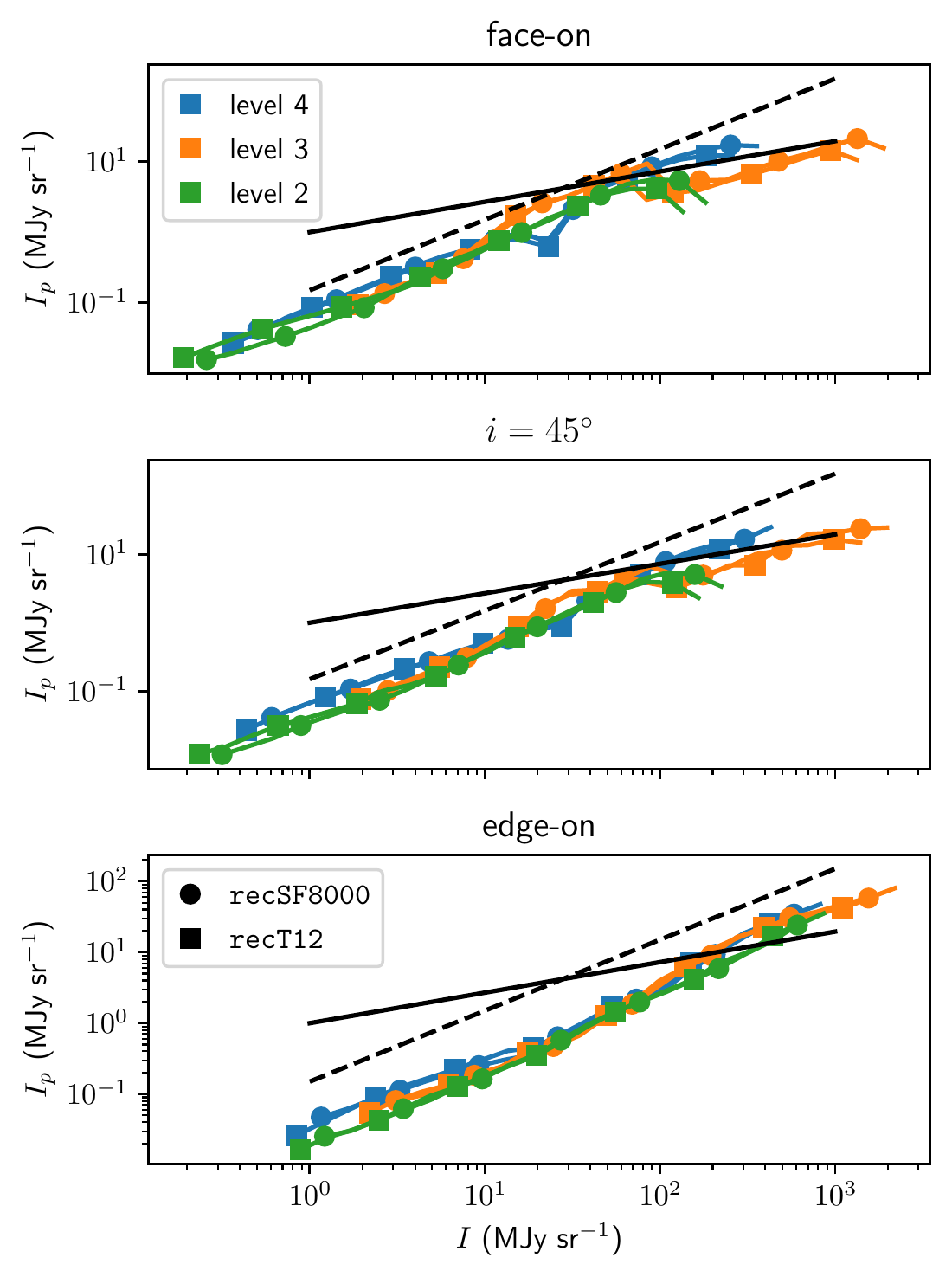}
    \caption{Same as \figureref{fig:production_I_vs_Ip}, but now showing the polarised intensity as a function of total intensity for the 3 different resolutions of the Auriga 6 galaxy.}
    \label{fig:resolution_I_vs_Ip}
\end{figure}

Given the importance of line-of-sight averaging for the synthetic polarisation images presented in this work, it is important to check how robust our results are against changes in resolution of the base model that defines the dust geometry. To this end, we compared synthetic images for three available resolution levels of the Auriga 6 galaxy: the fiducial level 4 simulation, the higher resolution level 3 simulation (used throughout this work), and a level 2 simulation with even higher resolution. It is important to note that the latter model shows some clear differences with the level 3 and 4 Auriga 6 galaxy: its dust distribution is generally more compact and peaks at lower dust densities. We therefore do not expect to see strict convergence of all our results.

\figureref{fig:resolution_histograms} shows the average histograms for all observer positions in the Auriga 6 all-sky map, for the 3 different model resolutions. The lowest resolution level 4 result on average has the highest linear polarisation fraction. The level 2 and 3 simulations show more compatible histograms, but still show a lot of scatter. The systematically higher polarisation fraction in the level 4 simulation is indicative of insufficient line-of-sight averaging. The scatter in the higher resolution results hints at the geometrical differences in these models, while the lack of clear systematic differences shows that the level 3 simulation has sufficient resolution for our purpose. The polarisation angle dispersion function shows a clear increase with resolution when large sky fractions are considered. This is mainly caused by a significantly better sky coverage in the high resolution models. When only considering the thin disc, the level 2 and 3 simulation again show reasonable agreement.

In \multifigref{fig:resolution_pstats}{fig:resolution_I_vs_Ip} we investigate how our nearby galaxy predictions change with resolution. This is less trivial to assess, since the three different resolution Auriga 6 galaxies have a different morphology, making them almost look like three different galaxies. As a result, the intensity-weighted average linear polarisation fraction and maximum polarised intensity are quite different between the different resolutions, with the highest resolution galaxy on average having the lowest linear polarisation fraction. This could indicate a lack of line-of-sight averaging in the level 3 and 4 simulations. The level 2 simulation however also fails to reproduce the highest intensities found in the level 3 simulation, showing that the level 2 images do not contain the bright, dense regions found in the other models. This indicates a clear morphological difference that could also explain the discrepancies we find. Overall, it looks like the $I-I_p$ relation is consistent between the different resolutions, while the polarisation maps fall within the scatter found for the level 3 models. Our nearby galaxy results are less sensitive to resolution than our all-sky results, and we do not expect these results to change significantly if we were to use the level 2 or level 4 simulations.

\end{appendix}

\end{document}